\documentclass[aps,prd,reprint,floatfix,nofootinbib,superscriptaddress]{revtex4-2}

\usepackage{graphicx}
\usepackage{subfigure}
\usepackage{amsmath,amssymb,amsfonts}
\usepackage{mathtools}
\usepackage{breakcites}

\begin{document}

\title{Holographic Butterfly Velocity  and Heavy-Quark Drag  in a Doubly Anisotropic Magnetized QGP}

\author{Le Zhang}
\email{zlphys@csuc.edu.cn}
\affiliation{School of General Education, Cyberspace Security University of China, Wuhan, 430040, China}
\affiliation{College of Physics and Electronic Science, Hubei Normal University, Huangshi, 435002, China}

\author{Xiao-Wei Zhang}
\affiliation{DUT-BSU Joint Institute, Dalian University of Technology, Dalian 116024, China}
\affiliation{College of Physics and Electronic Science, Hubei Normal University, Huangshi, 435002, China}

\author{Guo-Dong Zhou}
\affiliation{South-Western Institute for Astronomy Research, Yunnan University, Kunming 650500, China}
\affiliation{College of Physics and Electronic Science, Hubei Normal University, Huangshi, 435002, China}

\date{\today}

\begin{abstract}
We investigate butterfly velocity and heavy-quark energy loss in a strongly 
coupled quark-gluon plasma with two distinct sources of anisotropy, 
generated by collision geometry and an external magnetic field, within a 
five-dimensional holographic model exhibiting magnetic catalysis. 
In contrast to studies restricted to isotropic or singly anisotropic media, 
we analyze the butterfly velocity and heavy-quark drag force along all three 
orthogonal spatial directions at finite temperature and chemical potential in the deconfined phase. We find that the collision-induced anisotropy plays a dominant role in the directional dependence of both butterfly velocity and heavy-quark drag. It governs the directional splitting of the butterfly velocity, and significantly enhances the parallel drag force while suppressing the transverse components. For butterfly velocity, a strong magnetic field primarily suppresses the longitudinal butterfly velocity while leaving the transverse components nearly unchanged. For the drag force, the magnetic field breaks the transverse symmetry, yet it does not alter the overall monotonic increase of the drag force with heavy-quark velocity, temperature, and chemical potential.
Most importantly, by bridging the Einstein and string frames via the Weyl transformation, we extract the frame-dependent drag–chaos relation. Generalizing it to the doubly anisotropic plasma, we demonstrate that despite the underlying frame-dependent chaos, the normalized directional drag exhibits a direction-dependent geometric correction at order $\mathcal{O}(v^{3})$. These results establish a  direction-resolved connection between heavy-quark dissipation, information scrambling, and the underlying anisotropic geometry of magnetized QGP.
\end{abstract}

\maketitle

\section{Introduction}
High-energy heavy-ion collisions at the Large Hadron Collider (LHC) and the Relativistic Heavy-Ion Collider (RHIC) create a state of deconfined quarks and gluons known as the quark--gluon plasma (QGP)~\cite{BRAHMS:2004adc, Shuryak:2004cy, STAR:2005gfr, Gyulassy:2004zy}. experimental observations indicate that the QGP is strongly coupled~\cite{STAR:2005gfr, PHENIX:2004vcz, Gyulassy:2004zy, Busza:2018rrf, ALICE:2022wpn}, with dynamics governed by nonperturbative effects~\cite{Baier:1996kr, Eskola:2004cr}. Lattice quantum chromodynamics (lattice QCD), the principal nonperturbative method for studying strong interactions, has achieved substantial progress (see, e.g., Ref.~\cite{Meyer:2007ic}), but remains challenging for real-time nonequilibrium phenomena. The AdS/CFT correspondence~\cite{Maldacena:1997re, Gubser:1998bc, Witten:1998qj} therefore provides a promising nonperturbative framework for investigating strongly coupled non-Abelian plasmas (see Refs.~\cite{Casalderrey-Solana:2011dxg, DeWolfe:2013cua} for reviews of applications to the QGP).
In particular, gauge/gravity duality has been widely applied to theoretical and phenomenological studies of jet quenching in strongly coupled plasmas, including the drag force~\cite{Gubser:2006bz, Herzog:2006gh}, diffusion coefficients~\cite{Casalderrey-Solana:2006fio, Gubser:2006qh}, the jet-quenching parameter~\cite{Liu:2006ug, Liu:2006he}, the instantaneous energy loss of light quarks~\cite{Chesler:2008uy,Chesler:2008wd, Ficnar:2012np, Ficnar:2013wba, Ficnar:2013qxa}, and modifications of jet shapes and jet evolution~\cite{Hatta:2008tx, Chesler:2008hg}. 

As a many-body system containing a large number of deconfined quarks and gluons, the QGP also provides a setting in which to explore quantum-chaotic phenomena. An important quantity characterizing quantum chaos in strongly coupled many-body systems is the butterfly velocity, which may be interpreted as an effective speed of information propagation~\cite{Nomura:2017grg}. In gauge/gravity duality, black holes in asymptotically anti-de Sitter (AdS) spacetime are dual to thermal states of strongly coupled many-body quantum systems, allowing the butterfly velocity to be studied holographically in general spacetimes~\cite{Nomura:2017grg, Jahnke:2018off}. The butterfly velocity can be computed holographically using three methods that yield the same result. The first is shock wave analysis, in which out-of-time-order correlators (OTOCs) diagnose quantum chaos and correspond, on the gravity side, to near-horizon shock wave collisions~\cite{Shenker:2013pqa, Shenker:2013yza}. The second is pole-skipping analysis, which extracts the butterfly velocity from the pole-skipping behavior of the energy-density two-point function~\cite{Blake:2017ris, Blake:2018leo, Gursoy:2020kjd, Ahn:2020baf}. The third is the entanglement-wedge approach, which determines the butterfly velocity from the evolution of the entanglement wedge~\cite{Mezei:2016zxg, Grozdanov:2017ajz}. Recent studies have further related transport properties of strongly coupled plasmas to quantum chaos, including charge diffusion~\cite{Blake:2016wvh}, thermal diffusion~\cite{Blake:2017qgd}, the drag force, and the jet-quenching parameter~\cite{Ageev:2021poy}. Moreover, recent studies have shown that anisotropy of the medium can lead to frame-dependent chaos~\cite{Shukla:2023pbp, Shukla:2024qlf, Lee:2025ias, Nongmaithem:2026rop}, which means quantum-chaotic phenomena can differ between the Einstein and string frames. This complicates the relation between transport coefficients and quantum chaos in anisotropic media.

The five-dimensional Einstein-Maxwell-Dilaton (EMD) model, as a phenomenological gravity dual for hot and dense QCD matter, was proposed in Ref.~\cite {DeWolfe:2010he}.
In the black hole background, the EMD model is mapped onto the thermodynamic properties of QGP in the boundary field theory through the Hawking temperature and chemical potential of the  $AdS_5$ black hole solutions.
This EMD model has opened up a fruitful research area in the holographic application to the strongly coupled QGP.  This EMD model has found widespread application in the QCD phase diagram and critical end point~\cite{DeWolfe:2010he, He:2013qq, Yang:2014bqa, Yang:2015aia, Dudal:2018ztm, Chen:2018vty, Chen:2020ath, Chen:2019rez, Knaute:2017opk, Grefa:2021qvt, Cai:2022omk, Li:2023mpv, Li:2025lmp, Rougemont:2023gfz, Zhao:2023gur, Fu:2024wkn, Jokela:2024xgz, Chen:2024edy, Chen:2023yug, Chen:2024mmd, Chen:2024ckb, Zhu:2025gxo, Shen:2025yrn,  Liu:2026rmo},  heavy quark potential \cite{Chen:2021gop, Zhou:2020ssi}, heavy quarkonium~\cite{Chang:2024ksq, Zhou:2020ssi, Zhu:2025xiz},  transport properties \cite{DeWolfe:2011ts, Rougemont:2017tlu, Grefa:2022sav, Chen:2025fpd, Villarreal:2025ydo}   and energy loss of heavy and light quarks~ \cite{Rougemont:2015wca, Grefa:2022sav, Zhu:2023aaq, Chen:2024epd, Li:2025ugv, Chen:2025fpd, Zhang:2025wxi, Zhu:2025edv, Khan:2026uqc, Arefeva:2026yms}. Furthermore, EMD models have proven highly effective in investigating the quantum-chaotic phenomena of QCD matter~\cite{Shukla:2023pbp, Shukla:2024qlf, Lee:2025ias, Nongmaithem:2026rop}.

It is well established that the QGP immediately after heavy-ion collisions (HIC) is an anisotropic medium, exhibiting two types of anisotropy: (i) longitudinal-transverse anisotropy~\cite{Strickland:2013uga, ALICE:2015juo} and (ii) transverse-plane anisotropy induced by the magnetic field in non-central ion collisions~\cite{Skokov:2009qp, Voronyuk:2011jd}. Consequently, the EMD model is modified to holographically describe this doubly anisotropic magnetized QGP. Currently, EMD models describing this QGP include the light-quark holographic model with inverse magnetic catalysis~\cite{Arefeva:2022bhx, Arefeva:2022avn}, the heavy-quark holographic model with inverse magnetic catalysis~\cite{Arefeva:2020vae}, and the heavy-quark holographic model with magnetic catalysis~\cite{Arefeva:2023jjh}. In particular, the model of Ref.~\cite{Arefeva:2023jjh} has been shown to describe the QCD phase diagram with magnetic catalysis for heavy quarks in the doubly anisotropic magnetized QGP. As revealed by lattice QCD, the magnetic field's impact on the QCD phase diagram is quark-mass dependent: inverse magnetic catalysis occurs for light quarks, whereas magnetic catalysis takes place for heavy quarks. 

While this doubly anisotropic magnetized holographic model has been established as a description of the heavy-quark phase diagram with magnetic catalysis, its implications for quantum chaos and energy loss have not yet been explored. In this work, we adopt the doubly anisotropic heavy-quark holographic model with magnetic catalysis proposed in Ref.~\cite{Arefeva:2023jjh} as a phenomenological background, and we investigate the butterfly velocity,  the drag force acting on a heavy-quark probe, and the relation between them in this background. We find that the butterfly velocity is frame-dependent and derive analytic expressions for it in both the Einstein and string frames. Since the butterfly velocity is computed in the Einstein frame via shock waves and the drag force is obtained from trailing-string dynamics in the string frame, we bridge the two frames via the Weyl transformation to establish a consistent relation between the drag force and the frame-dependent butterfly velocity. After normalizing the directional drag force to isolate the overall frame factor, we find that the residual direction dependence appears as a geometric correction at order  $\mathcal{O}(v^{3})$,
where $v$ is the heavy-quark velocity.

The remainder of the paper is organized as follows. In Sec.~\ref{sec:EMD}, we briefly review the doubly anisotropic five-dimensional holographic model supported by an Einstein--dilaton--three-Maxwell action and proposed in Ref.~\cite{Arefeva:2023jjh}. In Sec.~\ref{secBV}, we calculate the butterfly velocities via shock waves in the Einstein frame and study their anisotropic behavior. In Sec.~\ref{sec:LOSS}, we derive the drag force on a heavy-quark probe moving through this background and analyze its directional dependence numerically. In Sec.~\ref{SEC:origin}, after bridging the Einstein and string frames via the Weyl transformation, we examine the frame-dependent relation between the drag force and the butterfly velocity and discuss the direction-dependent geometric correction to the normalized directional drag. Our conclusions are presented in Sec.~\ref{sec:CON}.

\section{The gravity background}
\label{sec:EMD}
In this section, we review the doubly anisotropic holographic background obtained in Ref.~\cite{Arefeva:2023jjh}. As described there, the model reproduces characteristic features of the heavy-quark phase diagram and the magnetic-catalysis behavior expected from lattice QCD. The Lagrangian in the Einstein frame is used in Refs.~ \cite{Arefeva:2020vae, Arefeva:2023jjh}
\begin{widetext}
\begin{eqnarray}
  {\cal L} = \sqrt{-g} \left[ R
    - \cfrac{f_0(\phi)}{4} \, F_0^2
    - \cfrac{f_1(\phi)}{4} \, F_1^2
    - \cfrac{f_3(\phi)}{4} \, F_3^2
    - \cfrac{1}{2} \, \partial_{\mu} \phi \, \partial^{\mu} \phi
    - V(\phi) \right]. \label{eq:2.01}
\end{eqnarray}
\end{widetext}
Where $\phi= \phi(z)$ is the scalar field and $V(\phi)$ is
the scalar field potential. The coupling functions $f_0(\phi)$,
$f_1(\phi)$ and $f_3(\phi)$ are associated with
the field strengths $F_0$, $F_1$ and $F_3$ of Maxwell fields, which are regarded as first, second, and third Maxwell fields, respectively.

To consider the Lagrangian~(\ref{eq:2.01}), the metric ansatz in the Einstein frame is given by
\begin{widetext}
\begin{eqnarray}
d s^2 &=& \frac{L^2}{z^2}\mathfrak{b}(z)\left[-g(z)dt^2 + dx_1^2 + \left(\frac{z}{L}\right)^{2-\frac{2}{\nu}} dx_2^2 + e^{c_B z^2}\left(\frac{z}{L}\right)^{2-\frac{2}{\nu}} dx_3^2 + \frac{dz^2}{g(z)}\right],\;\; \mathfrak{b}(z)=e^{2 {\cal A} (z)}.
\label{metricansatz}
\end{eqnarray}
\end{widetext}
Here, $L$ denotes the $AdS_5$ radius and is conventionally set to unity, $L=1 \,\text{GeV}^{-1}$. The parameter $\nu$ characterizes the primary anisotropy induced by the asymmetry of a heavy-ion collision, whereas $c_B$ characterizes the secondary anisotropy associated with the external magnetic field $F_{3}=q_3\,d x^1 \wedge d x^2$, where $q_3$ is a constant magnetic ``charge.'' For $\nu=1$, the longitudinal-transverse anisotropy disappears, leaving only the magnetic-field-induced anisotropy, whereas for $c_B=0$, it reduces to the zero-magnetic-field case.
The function $\mathfrak{b}(z)$ is the warp factor determined by ${{\cal A} (z)}$. $g(z)$ is the blackening function, with the usual boundary conditions $g(0)=1$ and $g(z_h)=0$. Here, $z_h$ denotes denotes the location of the black-hole horizon and determines the Hawking temperature. 

Following Ref.~\cite{Arefeva:2023jjh}, magnetic catalysis is realized; for sufficiently small $c_B$ the warp factor can be written as,
\begin{eqnarray}
\mathfrak{b}(z)=e^{2 {\cal A} (z)}=e^{-c z^2/2-2(p-c_B q_3)z^4}, \label{warp factor1}
\end{eqnarray}
where, $c=4R_{gg}/3$, $R_{gg}=1.16\,\text{GeV}^2$, 
$p=0.273\,\text{GeV}^4$ ~\cite{Pirner:2009gr, He:2013qq, Li:2017tdz, Arefeva:2023jjh}. 
The corresponding blackening function derived in Ref.~\cite{Arefeva:2023jjh} is
\begin{eqnarray}
g(z)&=&e^{c_B z^2}\Bigg[1 - \frac{\tilde{I}_1(z)}{\tilde{I}_1(z_h)} + \frac{\mu^2(2R_{gg} + c_B(q_3-1))\tilde{I}_2(z)}{L^2\left(1 - e^{(2R_{gg}+c_B(q_3-1))\frac{z_h^2}{2}}\right)^2}\nonumber\\& &
\times \left(1 - \frac{\tilde{I}_1(z)\tilde{I}_2(z_h)}{\tilde{I}_1(z_h)\tilde{I}_2(z)}\right)\Bigg],
\label{blackening_func}
\end{eqnarray}
where
\begin{eqnarray}
\tilde{I}_1(z) &=& \int_0^z e^{(2R_{gg} - 3c_B)\frac{\xi^2}{2} + 3(p - c_B q_3)\xi^4}\xi^{1+\frac{2}{\nu}} d\xi \label{eq:I1}, \\
\tilde{I}_2(z) &=& \int_0^z e^{(2R_{gg} + c_B(q_3/2 - 2))\xi^2 + 3(p - c_B q_3)\xi^4}\xi^{1+\frac{2}{\nu}} d\xi \;\;\label{eq:I2}.
\end{eqnarray}
In Eq.~(\ref{blackening_func}), $\mu$ denotes the chemical potential. 

The Hawking temperature, evaluated at the horizon, is~\cite{Arefeva:2023jjh}
\begin{widetext}
\begin{eqnarray}
T &=& \frac{|g'|}{4 \pi}\Big|_{z=z_h}\nonumber\\
&=&\left|-\frac{e^{(2R_{gg}-c_B)\frac{z_h^2}{2} + 3(p-c_B q_3)z_h^4}z_h^{1+\frac{2}{\nu}}}{4\pi\tilde{I}_1(z_h)}
\times \left[1 - \frac{\mu^2(2R_{gg}+c_B(q_3-1))\left(e^{(2R_{gg}+c_B(q_3-1))\frac{z_h^2}{2}}\tilde{I}_1(z_h) - \tilde{I}_2(z_h)\right)}{L^2\left(1 - e^{(2R_{gg}+c_B(q_3-1))\frac{z_h^2}{2}}\right)^2}\right]\right|.
\label{temperature}
\end{eqnarray}
\end{widetext}

Following the procedure in Ref.~\cite{Arefeva:2023jjh} and using the solution for the warp factor in Eq.~\ref{warp factor1}, we can write the scalar field $\phi(z)$ as follows:
\begin{widetext}
\begin{eqnarray}
   \phi (z) &=& \int_{z_0}^z \Bigl[
    - \, 4 + \frac{2}{3} \, \nu \Biggl(
    6 + 3 \, (c_B \, (2 - 3 \nu) + 6 R_{gg} \nu) \, \xi^2
    + \bigl(- \, 3 \, c_B(c_B + 60 \, q_3) \, \nonumber \\
    & &+ 4 (45 \, p + R_{gg}^2) \bigr) \, \nu \xi^4
    + 48 R_{gg} (p - c_B \, q_3) \, \nu \xi^6
    + 144 \, (p-c_B \, q_3)^2 \, \nu \xi^8 \Bigr)
    \Biggr]^{\frac{1}{2}} \, \frac{d\xi}{\nu \xi}.
\label{eq:4.80}
\end{eqnarray}
\end{widetext}
The above formula implies the usual boundary condition, $\phi(z_0)=0$, where $z_0$ serves to fit the string tension behavior~\cite{Arefeva:2020byn}. In the present work, we set \(z_0=0.001 \text {GeV}^{-1}\), which is consistent with the parameters used in Ref.~\cite{Arefeva:2023jjh}.

%In the following, we apply this doubly anisotropic holographic framework to the butterfly velocity and drag force of a strongly coupled QGP exhibiting magnetic catalysis.

\section{Butterfly Velocity}
\label{secBV}
%Two established approaches are commonly used to compute the butterfly velocity in anisotropic backgrounds and yield consistent results. The first is a near-horizon shock wave calculation~\cite{Blake:2016wvh, Blake:2017qgd, Fischler:2018kwt, Jahnke:2017iwi}; the second relates the growth of holographic entanglement entropy to quantities characterizing quantum chaos~\cite{Gursoy:2020kjd}. For a review, see Ref.~\cite{Jahnke:2018off} and references therein. 

In this section, we briefly present the shock wave calculation of the butterfly velocity in a doubly anisotropic magnetized QGP. 
For convenience, we now rewrite the Einstein frame metric, Eq.~(\ref{metricansatz}), for this doubly anisotropic holographic background as follows:
\begin{eqnarray}
d s_{(E)}^2 &=& G_{t\, t} dt^2 + \sum_{i =1}^{3} G_{x_i\, x_i} d x_i^2 + G_{z\, z}d z^2.
\label{Einstein metric}
\end{eqnarray}
where, $G_{t t}= -{z^{-2}}{\mathfrak{b}(z) g(z)}$,\; $G_{x_1\, x_1} ={z^{-2}}{\mathfrak{b}(z) }$,\; $G_{x_2\, x_2} = \mathfrak{b}(z) z^{-\frac{2}{\nu}}$,\; $G_{x_3\, x_3} =\mathfrak{b}(z) e^{c_B z^2} z^{-\frac{2}{\nu}}$, \;$G_{z\, z}= { {z^{-2}}\mathfrak{b}(z) }/{g(z)}$. $\mathfrak{b}(z) = e^{2{\cal A} (z)}$ denotes the warp factor in the Einstein frame. 
Here, the boundary of the bulk spacetime is set at $z=0$, and the horizon is located at $z=z_h$. It is assumed that at $z=z_h$,
$G_{tt} $ has a first order zero, $G_{z z} $ has a first order pole, and the other metric components are finite. Near the horizon, the metric functions 
$G_{tt} $ and  $G_{z z} $ can be approximated by a first-order zero and a first-order pole, respectively,
\begin{eqnarray}
G_{t t} = - c_0 (z-z_h),\;\;\;  G_{z z} = \frac{c_1} {(z-z_h)}.
\label{Near the horizon}
\end{eqnarray}
Then the inverse Hawking temperature $\beta =1/T$ is derived from the condition that the Euclidean continuation of the above metric be regular at the horizon,
\begin{eqnarray}
\beta = 4 \pi \sqrt{\frac{c_1}{c_0}}.
\end{eqnarray}

To employ the shock wave calculation approach~\cite{Jahnke:2017iwi}, we need to introduce the Tortoise coordinate $z_{*}$ as
\begin{eqnarray}
d z_{*}= -\sqrt{\frac{G_{zz}}{- G_{t t}}} d z,
\end{eqnarray}
along with defining the Kruskal coordinates $U$ and $V$
\begin{eqnarray}
U V = -e^{\frac{4\pi z_*}{\beta}},\;\;\; \frac{U}{V} = -e^{-\frac{4\pi t}{\beta}}.
\end{eqnarray}
In terms of these coordinates, the metric in the Einstein frame, Eq.~(\ref{Einstein metric}), can be rewritten as
\begin{eqnarray}
\label{unperturbed metric}
& &ds_{(E)}^2 = 2 A(U, V) dU dV + \sum_{i=1}^{3}G_{x_i x_i}(U, V)d  x_i^2,\;\;\;
\\
& &A(U, V) = -\frac{\beta^2 G_{tt}}{8 \pi^2 } \frac{1}{U V}.
\end{eqnarray} 
In these coordinates, the horizon lies at $U=0$ or $V=0$, the boundary lies at $U V = -1$, and the black hole singularity is at $U V = 1$.

Following the shock wave procedure~\cite{Sfetsos:1994xa, Arefeva:2016upm, Jahnke:2017iwi}, a null energy pulse propagating at the speed of light along the \(V\)-direction is introduced at \(U=0\). The null shock front at \(U=0\) splits the spacetime into two regions, the left region (L) and the right region (R). Region R (\(U<0\)) lies on one side of the shock front, and its metric remains consistent with the unperturbed metric (Eq.~\ref{unperturbed metric}); region L (\(U>0\)) lies on the other side of the shock front, and its metric must be modified to incorporate the influence of this energy pulse.

To incorporate the backreaction of this energy pulse in anisotropic backgrounds, we can modify the unperturbed metric through $V \rightarrow V + \theta(U) \alpha(t, x_i)$, where $\alpha(t, x_i )$ is a function that can be determined from Einstein's equations~\cite{Sfetsos:1994xa, Dray:1984ha}. 

With the logic above, the Einstein frame metric can be written as
\begin{widetext}
\begin{eqnarray}
d s_{(E)}^2 = 2 \hat{A}(\hat{U}, \hat{V}) d \hat{U} d \hat{V}  + \sum_{i} G_{x_i x_i}(\hat{U}, \hat{V}) d \hat{x}_i^2 - 2\hat{A}(\hat{U}, \hat{V}) \hat{\alpha}(t, \hat{x}_i) \delta(\hat{U}) d\hat{U}^2,
\end{eqnarray}
\end{widetext}
here, for convenience, we have defined the new coordinates as follows:
\begin{eqnarray}
\hat{U} = U,\;\; \hat{V}= V +\theta(U) \alpha(t, x_i),\;\;\hat{x}_i =x_i.
\end{eqnarray}
The stress-energy tensor of the shock wave can be assumed to take the form:
\begin{eqnarray}
T_{shock} = E  e^{2\pi t/\beta}\delta(\hat{U})a(\hat{x}_i)d\hat{U}^2,
\end{eqnarray}
where $E$ is a constant parameter for the pulse's asymptotic energy and $a(\hat{x})$ specifies the spatial profile of the perturbation. Within anisotropic backgrounds where the perturbation is taken to be localized, the function is specified as $a(x_i) = \delta (\hat{x}_i)$. 

Following the Einstein equation solution procedure (further computational details can be found in Ref.~\cite{Jahnke:2017iwi}),the equation for \(\hat{\alpha}\) at the horizon \(U=0\) can be solved for a homogeneous perturbation (\(a = \text{constant}\)) by making the ansatz \(\hat{\alpha} = \text{constant} \times \exp(2\pi t/\beta)\), confirming that the Lyapunov exponent saturates the chaos bound. 

In the case of localized perturbations, we consider the perturbation propagating along a specific spatial direction \(x_i\) (no sum over \(i\)), such that \(a(x_i) = \delta(x_i)\). This leads to an equation for \(\hat{\alpha}(t, x_i)\) of the form:
\begin{eqnarray}
\label{eqforalpha}
\big(\partial_{{x}_i}^2-\mu_{i}^2\big) {\hat{\alpha}} (t,{x}_i)=\frac{G_{{x}_i {x}_i}(z_{\text{h}})}{\hat{A}(z_{\text{h}})}\,8\pi E\,e^{2\pi t/\beta}\,\delta({x}_i),
\end{eqnarray}
with
\begin{eqnarray}
\mu_i^2 = \frac{z_h {G_{x_i x_i}(z_h)}}{2\hat{A}(z_h)} \displaystyle\sum_{j=1}^3\frac{G_{x_j x_j}'(z_h)}{G_{x_j x_j}(z_h)}.
\end{eqnarray}
By evaluating \(G_{tt}\) with the near-horizon expression and imposing \(\hat{A}(z_h)= 2 c_1 z_h\), the preceding equation admits a further simplification, and the solution of Eq.~(\ref{eqforalpha}) for large \(x_i\) (taking \(x_i > 0\)) is given by
\begin{eqnarray}
{\alpha}(t, x_i) \sim \exp\left[\frac{2\pi}{\beta} (t-t_*) - \mu_i x_i\right].
\end{eqnarray}
Such asymptotic behavior indicates that the Lyapunov exponent saturates the chaos bound $\lambda_L = 2\pi/\beta$, and that the distinct butterfly velocities $v_{B,{(i)}}$ along each spatial direction $x_i$ are determined by the corresponding coefficient $\mu_i$,
\begin{eqnarray}
v_{B,{(i)}} = \left(\frac{2\pi }{\beta \mu_i}\right) = \sqrt{\frac{-G^{x_i x_i}(z_h)G_{t t}'(z_h)}{\displaystyle\sum_{j=1}^3 G^{x_j x_j}(z_h) G_{x_j x_j}'(z_h)}}.
\label{vbappendix}
\end{eqnarray}
This expression coincides with the standard shock wave result for a diagonal anisotropic geometry~\cite{Blake:2016wvh,Blake:2017qgd,Fischler:2018kwt,Jahnke:2017iwi}. In that construction, the near-horizon behavior is given in Eq.(\ref{Near the horizon}).  Substituting the metric in Eq.~(\ref{Einstein metric}) gives the directional butterfly velocities in the doubly anisotropic QGP:
\begin{eqnarray}
v_{B,{(1)}} &=& \sqrt{\frac{4\ \pi\ z_h\ T}{\frac{2\,(2+\nu)}{\nu} - 2\ c_B\ z_h^2 - \frac{3\ z_h\ \mathfrak{b}'(z_h)}{\mathfrak{b}(z_h)}}},\label{33vb1}\\
v_{B,{(2)}}&=& \sqrt{\frac{4\ \pi\ T\ z_h^{-1+\frac{2}{\nu}}}{\frac{2\,(2+\nu)}{\nu} - 2\ c_B\ z_h^2 - \frac{3\ z_h\ \mathfrak{b}'(z_h)}{\mathfrak{b}(z_h)}}},\label{34vb2}\\
v_{B,{(3)}} &=& \sqrt{\frac{4\ \pi\ T\ z_h^{-1+\frac{2}{\nu}}\ e^{-c_B z_h^2}\ }{\frac{2\,(2+\nu)}{\nu} - 2\ c_B\ z_h^2 - \frac{3\ z_h\ \mathfrak{b}'(z_h)}{\mathfrak{b}(z_h)}}}.\label{35vb3}
\end{eqnarray}
Here, the Hawking temperature definition in Eq.~(\ref{temperature}) is substituted.

From Eqs.~(\ref{33vb1})–(\ref{35vb3}), we readily obtain the following relation.
\begin{eqnarray}
\label{vbrelationship}
v_{B, (1)} = v_{B, (2)} \cdot z_h^{1-\frac{1}{\nu}} = v_{B, (3)} \cdot z_h^{1-\frac{1}{\nu}} \cdot e^{\frac{1}{2} c_B z_h^2}.
\end{eqnarray}

The isotropic result is recovered by setting $\nu=1$ and $c_B=0$:
\begin{eqnarray}
v_{B,{(iso)}} = \sqrt{\frac{- G_{tt}'(z_h)  }{3\ G_{x_1 x_1}'(z_h)}}=\sqrt{\frac{4\ \pi\ z_h\ T}{ 6 - \frac{3\ z_h\ \mathfrak{b}'(z_h)}{\mathfrak{b}(z_h)}}}.
\label{vbiso}
\end{eqnarray}

We now discuss the numerical behavior of the butterfly velocities in a strongly coupled doubly anisotropic QGP. All calculations are performed in the pure deconfined QGP phase above the phase-transition point identified from the data of Ref.~\cite{Arefeva:2023jjh}. The phenomena discussed below therefore characterize the deconfined plasma rather than the critical behavior near the phase transition.

\begin{figure*}[t]
\centering
 \subfigure[$v_{B, ({iso})}$ vs. $T$]{
\label{figvb1a}
\includegraphics[width=0.40\linewidth]{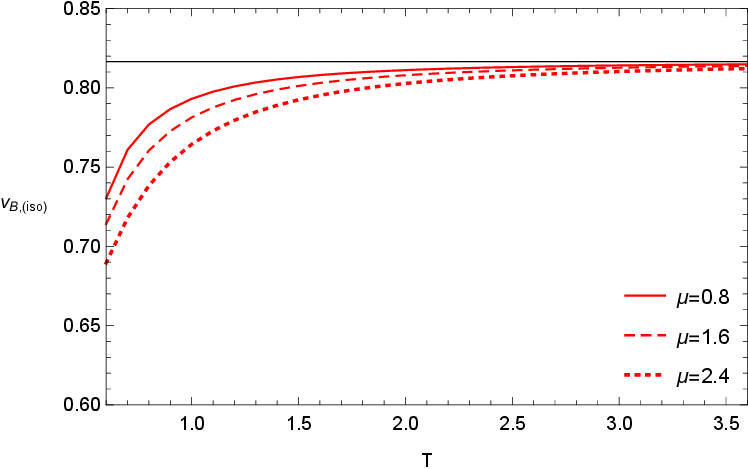}}
\subfigure[$v_{B, ({iso})}$ vs. $\mu$]{
\label{figvb1b}
\includegraphics[width=0.40\linewidth]{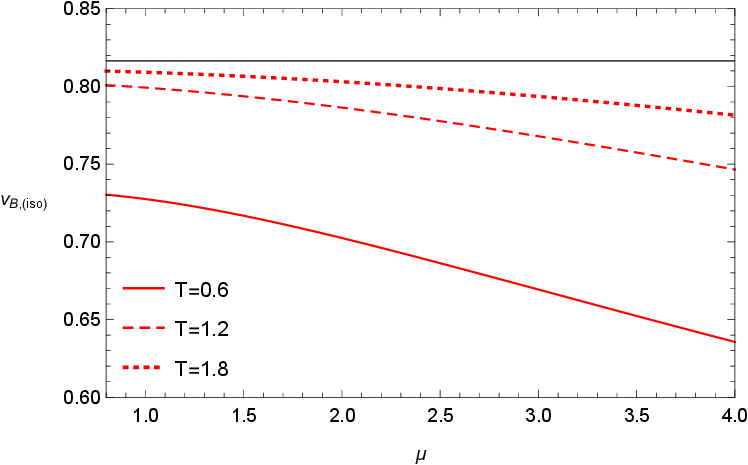}}
\caption{Isotropic butterfly velocity $v_B^{({iso})}$ as a function of (a) temperature $T$ for several values of $\mu$ and (b) chemical potential $\mu$ for several values of $T$. We use $R_{gg}=1.16\,\text{GeV}^2$, $p=0.273\,\text{GeV}^4$, and $q_3=5\,\text{GeV}^2$, with $[\mu]=[T]=\text{GeV}$.}
\label{figvb1}
\end{figure*}

Figure~\ref{figvb1} shows the temperature and chemical-potential dependence of the butterfly velocity in the isotropic, non-magnetized limit. The quantity $v_{B,({iso})}$ increases with temperature and approaches the conformal benchmark identified in Refs.~\cite{Giataganas:2017koz,Jahnke:2017iwi} at high $T$, while it decreases with chemical potential.

We next analyze the butterfly velocities in the doubly anisotropic QGP with magnetic catalysis and examine their dependence on the primary anisotropy, magnetic field, temperature, and chemical potential.
The holographic model constructed in Ref.~\cite{Arefeva:2023jjh} exhibits magnetic catalysis and reproduces characteristic features of the heavy-quark phase diagram.
That work obtained phase diagrams in the $(\mu,T)$ plane for both isotropic and anisotropic cases at several values of $c_B$ using a trial-and-error procedure.
In this setup, $x_1$ is the primary anisotropic direction, while $x_2$ and 
$x_3$ are the transverse directions; $x_3$ is the direction of the external magnetic field. The external magnetic field is characterized by a constant charge \(q_3\), with its metric coupling governed by the coefficient \(c_B\). Following the heavy-quark phase diagrams of Ref.~\cite{Arefeva:2023jjh}, we fix \(q_3=5\,\text{GeV}^2\), rendering \(c_B\) the only tunable magnetic parameter. In this model, the choice of \(c_B = -B^2\) is driven by considerations suggesting a decrease in the blackening function values toward the horizon~\cite{DHoker:2009ixq, Bohra:2019ebj, Arefeva:2022avn}. 
%So,  the magnetic-field strength increases with $|c_B|$ in our work.

\begin{figure*}[t]
\centering
\subfigure[$v_{B,(1)}$ vs. $\nu$ for $\mu=0.95$]{
\label{figvb2a}
\includegraphics[width=0.3\linewidth]{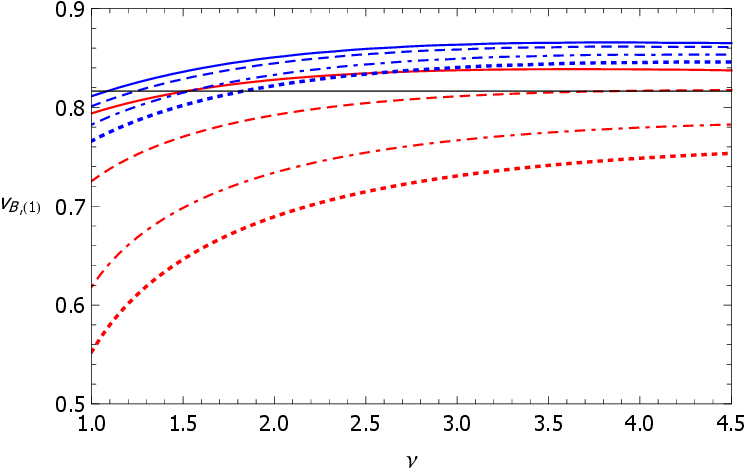}}
\subfigure[$v_{B,(2)}$ vs. $\nu$ for $\mu=0.95$]{
\label{figvb2b}
\includegraphics[width=0.3\linewidth]{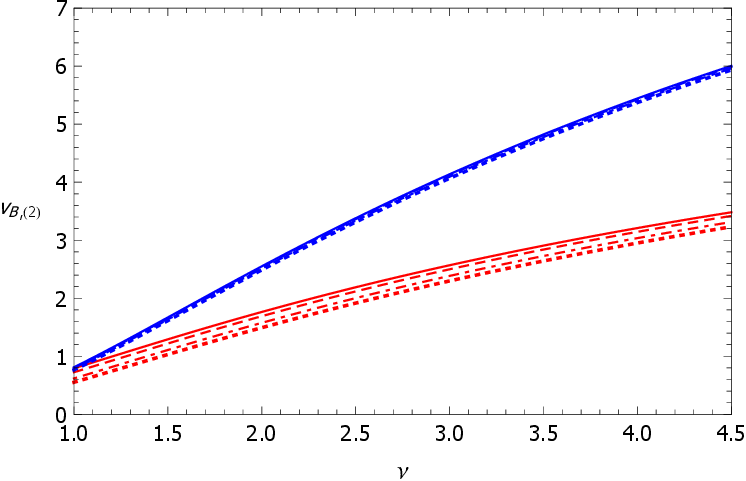}}
\subfigure[$v_{B,(3)}$ vs. $\nu$ for $\mu=0.95$]{
\label{figvb2c}
\includegraphics[width=0.3\linewidth]{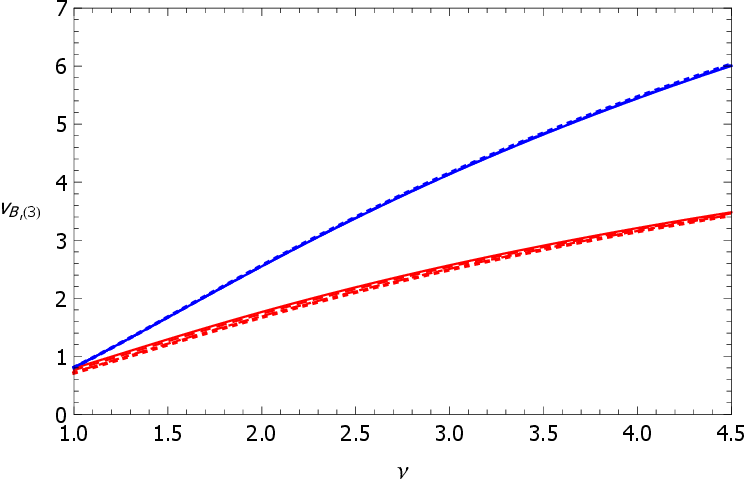}}
\subfigure[$v_{B,(1)}$ vs. $\nu$ for $\mu=1.9$]{
\label{figvb2d}
\includegraphics[width=0.3\linewidth]{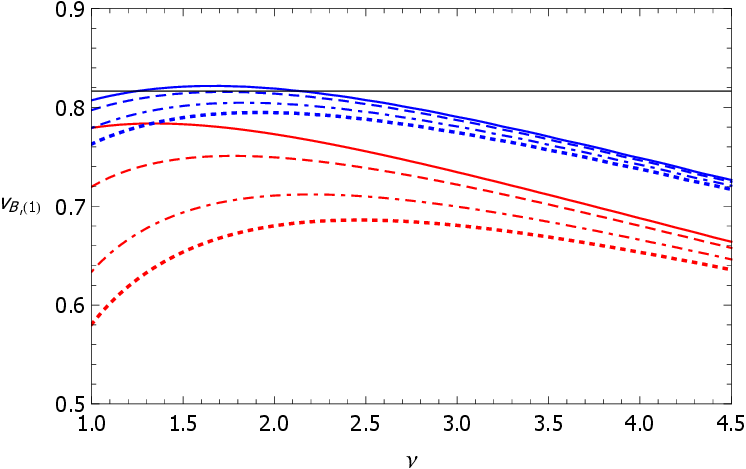}}
\subfigure[$v_{B,(2)}$ vs. $\nu$ for $\mu=1.9$]{
\label{figvb2e}
\includegraphics[width=0.3\linewidth]{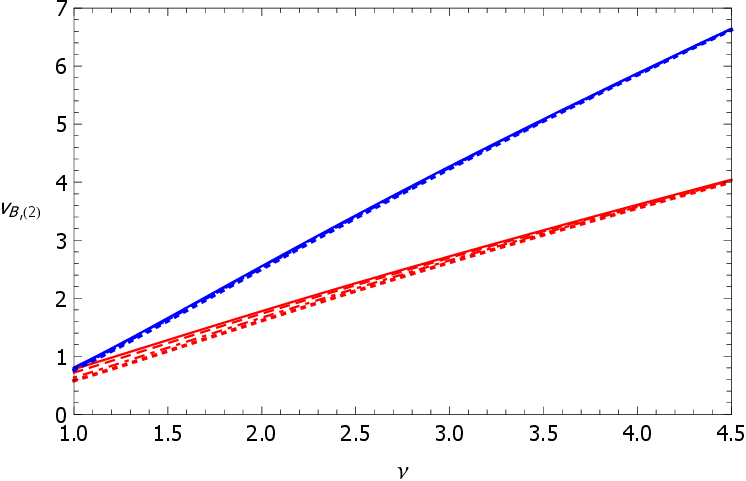}}
\subfigure[$v_{B,(3)}$ vs. $\nu$ for $\mu=1.9$]{
\label{figvb2f}
\includegraphics[width=0.3\linewidth]{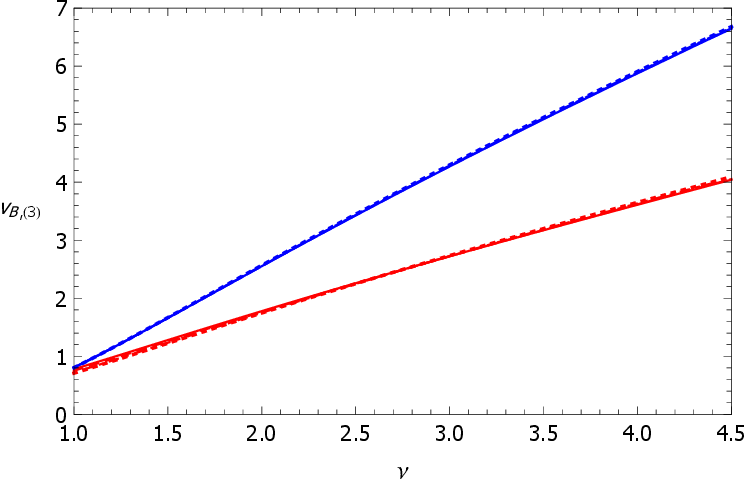}}
\caption{Butterfly velocities $v_{B, (i)}$ as functions of $\nu$. The first and second rows correspond to $\mu=0.95$ and $1.9$, respectively. The red and blue curves denote $T=1.05$ and $2.1$, while the solid, dashed, dot-dashed, and dotted curves correspond to $c_B=0,-1,-3$, and $-5$. We use $R_{gg}=1.16\,\text{GeV}^2$, $p=0.273\,\text{GeV}^4$, and $q_3=5\,\text{GeV}^2$, with $[\mu]=[T]=[c_B]^{1/2}=\text{GeV}$.}
\label{figvb2}
\end{figure*}

\begin{figure*}[t]
\centering
\subfigure[$v_{B,(1)}$ vs. $|c_B|$ for $\mu=0.95$]{
\label{figvb3a}
\includegraphics[width=0.3\linewidth]{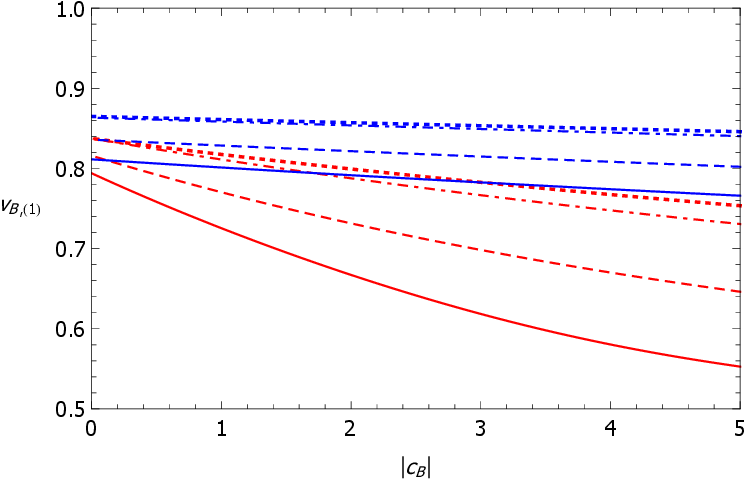}}
\subfigure[$v_{B,(2)}$ vs. $|c_B|$ for $\mu=0.95$]{
\label{figvb3b}
\includegraphics[width=0.3\linewidth]{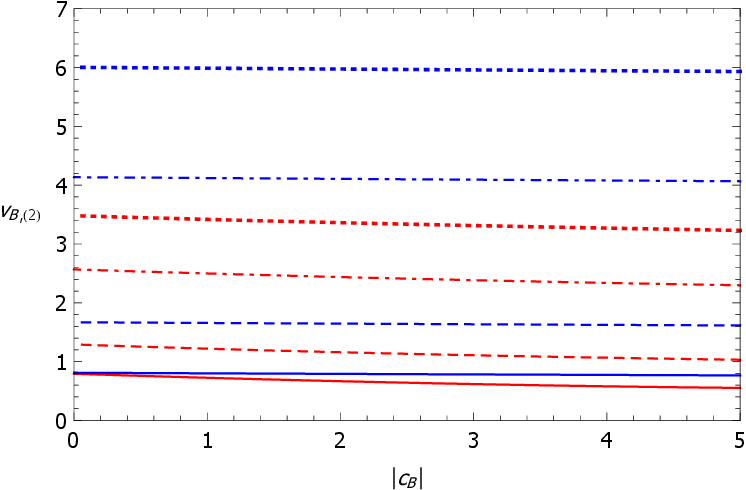}}
\subfigure[$v_{B,(3)}$ vs. $|c_B|$ for $\mu=0.95$]{
\label{figvb3c}
\includegraphics[width=0.3\linewidth]{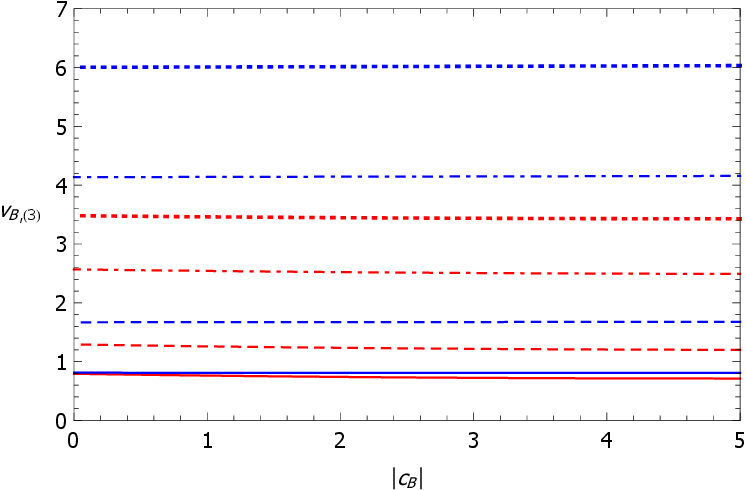}}
\subfigure[$v_{B,(1)}$ vs. $|c_B|$ for $\mu=1.9$]{
\label{figvb3d}
\includegraphics[width=0.3\linewidth]{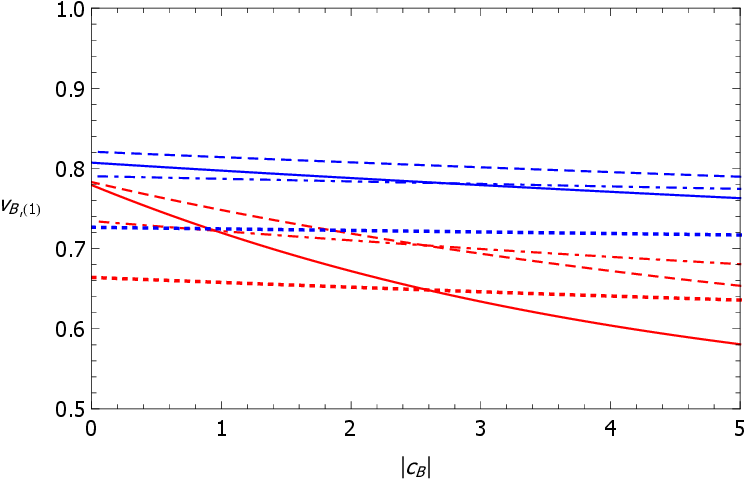}}
\subfigure[$v_{B,(2)}$ vs. $|c_B|$ for $\mu=1.9$]{
\label{figvb3e}
\includegraphics[width=0.3\linewidth]{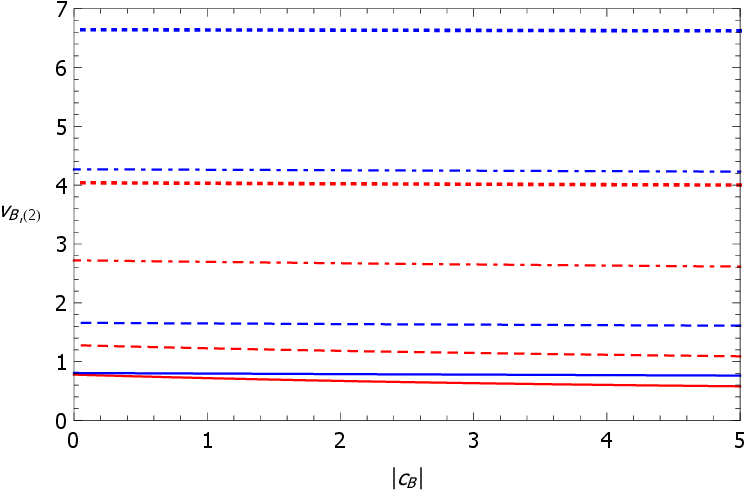}}
\subfigure[$v_{B,(3)}$ vs. $|c_B|$ for $\mu=1.9$]{
\label{figvb3f}
\includegraphics[width=0.3\linewidth]{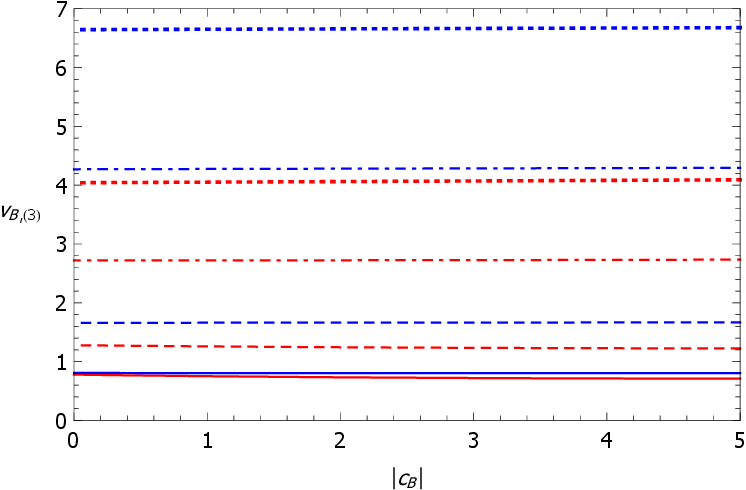}}
\caption{Butterfly velocities $v_{B,(i)}$ as functions of $|c_B|$. The first and second rows correspond to $\mu=0.95$ and $1.9$, respectively. The red and blue curves denote $T=1.05$ and $2.1$, while the solid, dashed, dot-dashed, and dotted curves correspond to $\nu=1,1.5, 3$, and $4.5$. We use $R_{gg}=1.16\,\text{GeV}^2$, $p=0.273\,\text{GeV}^4$, and $q_3=5\,\text{GeV}^2$, with $[\mu]=[T]=[c_B]^{1/2}=\text{GeV}$.}
\label{figvb3}
\end{figure*}

Figure~\ref{figvb2} shows that, for each fixed temperature and chemical potential,  the transverse components $v_{B,(2)}$ and $v_{B,(3)}$ increase monotonically with $\nu$. The dependence of the component along the primary anisotropic direction, $v_{B,(1)}$, on $\nu$ is more sensitively to the chemical potential: at low $\mu$ it increases and rapidly saturates, whereas at high $\mu$ it first increases and then decreases with $\nu$. Thus, at high chemical potential, this suppression of $v_{B, (1)}$ competes with and eventually overcomes the enhancement due to the anisotropy.
The primary anisotropy also produces a pronounced directional splitting, with the transverse butterfly velocities substantially larger than the longitudinal one. By contrast, the magnetic field generates only a small difference between the two transverse components. Figure~\ref{figvb3} further shows that increasing $|c_B|$ suppresses $v_{B,(1)}$, while leaving $v_{B,(2)}$ and $v_{B,(3)}$ nearly unchanged within the parameter range considered. The crossing of the curves in Figure~\ref{figvb3d} is consistent with the nonmonotonic dependence of $v_{B,(1)}$ on $\nu$ at sufficiently large chemical potential.
Compared with the isotropic case, both the longitudinal and transverse butterfly velocities can exceed the isotropic conformal benchmark, possibly because the metric used here lies outside the conformal class considered in Ref.~\cite{Gursoy:2020kjd}. In the coordinate normalization used here, the transverse components may also exceed unity. This observation has been discussed in Refs.~\cite{Balasubramanian:2019stt, Basu:2025exh, Basu:2026wzp} and interpreted as an apparent effect that does not imply a violation of relativistic causality, possibly indicating that the holographic doubly anisotropic QGP in this paper can be regarded as a strongly nonequilibrium holographic system.

\begin{figure*}[t]
\centering
\subfigure[$v_{B,(1)}$ vs. $T$ for $\mu=0.95$]{
\label{figvb4a}
\includegraphics[width=0.3\linewidth]{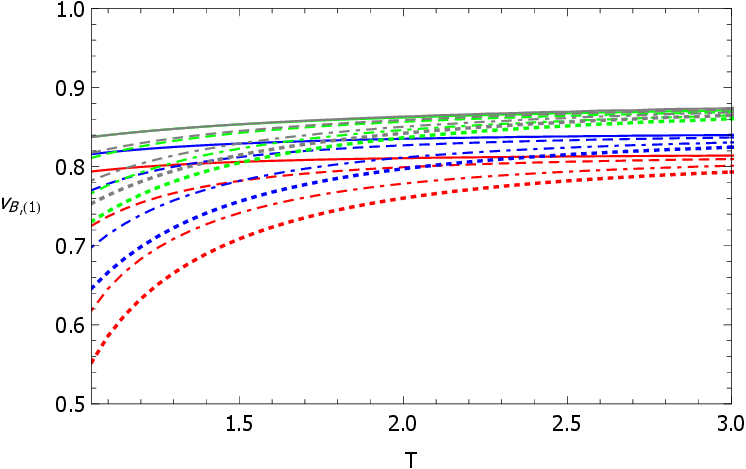}}
\subfigure[$v_{B,(2)}$ vs. $T$ for $\mu=0.95$]{
\label{figvb4b}
\includegraphics[width=0.3\linewidth]{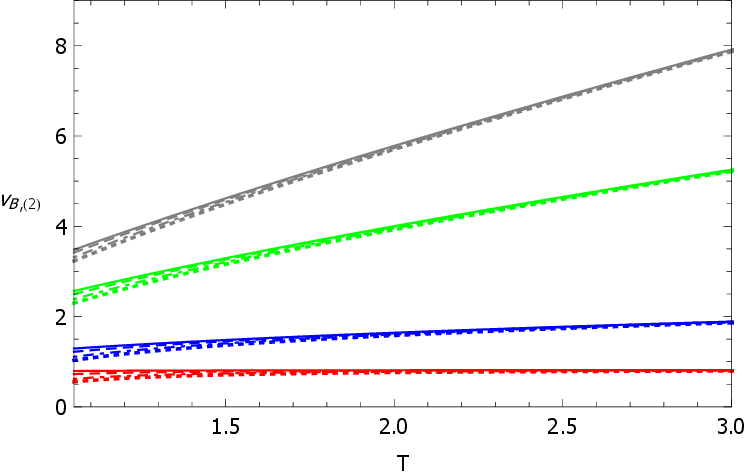}}
\subfigure[$v_{B,(3)}$ vs. $T$ for $\mu=0.95$]{
\label{figvb4c}
\includegraphics[width=0.3\linewidth]{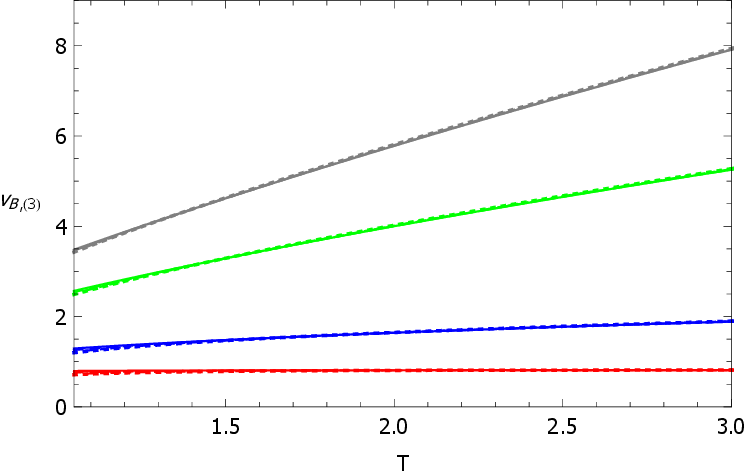}}
\subfigure[$v_{B,(1)}$ vs. $T$ for $\mu=1.9$]{
\label{figvb4d}
\includegraphics[width=0.3\linewidth]{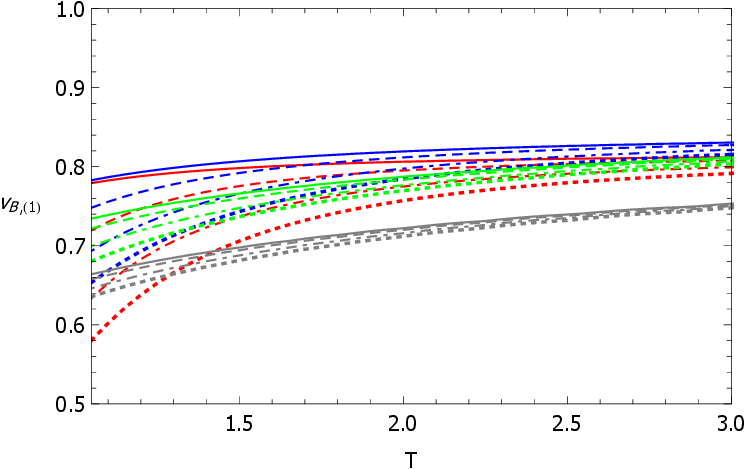}}
\subfigure[$v_{B,(2)}$ vs. $T$ for $\mu=1.9$]{
\label{figvb4e}
\includegraphics[width=0.3\linewidth]{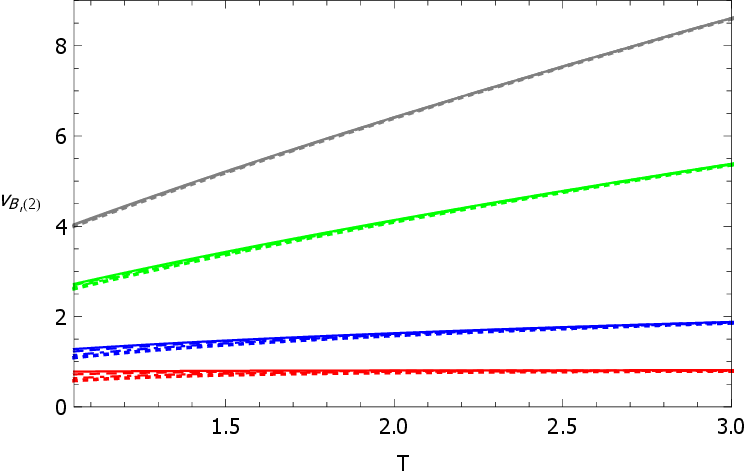}}
\subfigure[$v_{B,(3)}$ vs. $T$ for $\mu=1.9$]{
\label{figvb4f}
\includegraphics[width=0.3\linewidth]{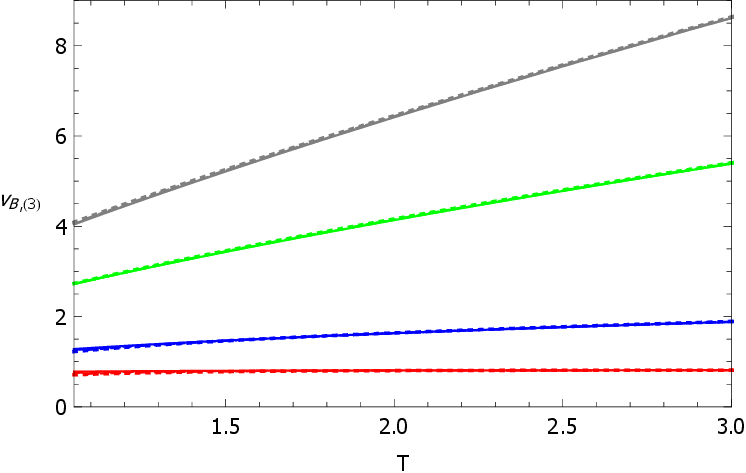}}
\caption{Butterfly velocities $v_{B,(i)}$ as functions of $T$. The first and second rows correspond to $\mu=0.95$ and $1.9$, respectively. The black, red, blue, and gray curves denote $\nu=1,1.5,3$, and $4.5$, while the solid, dashed, dot-dashed, and dotted curves correspond to $c_B=0,-1,-3$, and $-5$. We use $R_{gg}=1.16\,\text{GeV}^2$, $p=0.273\,\text{GeV}^4$, and $q_3=5\,\text{GeV}^2$, with $[\mu]=[T]=[c_B]^{1/2}=\text{GeV}$.}
\label{figvb4}
\end{figure*}

\begin{figure*}[t]
\centering
\subfigure[$v_{B,(1)}$ vs. $\mu$ for $T=1.05$]{
\label{figvb5a}
\includegraphics[width=0.3\linewidth]{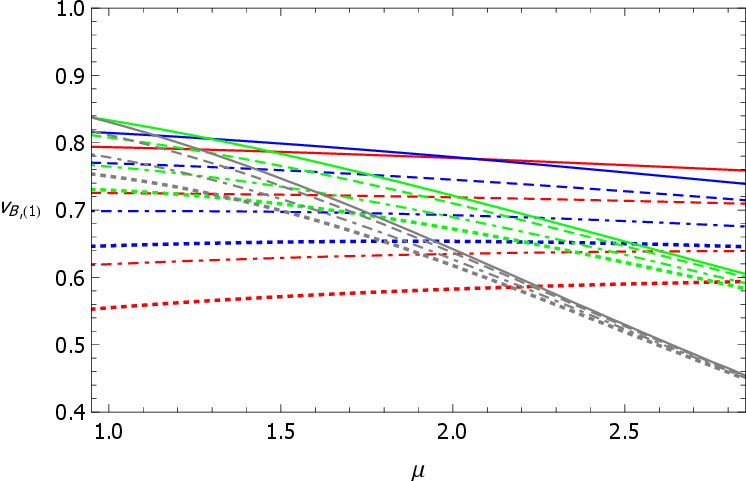}}
\subfigure[$v_{B,(2)}$ vs. $\mu$ for $T=1.05$]{
\label{figvb5b}
\includegraphics[width=0.3\linewidth]{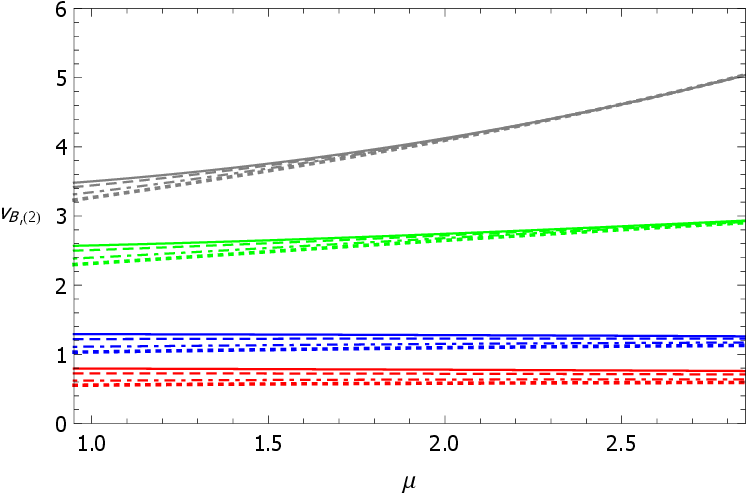}}
\subfigure[$v_{B,(3)}$ vs. $\mu$ for $T=1.05$]{
\label{figvb5c}
\includegraphics[width=0.3\linewidth]{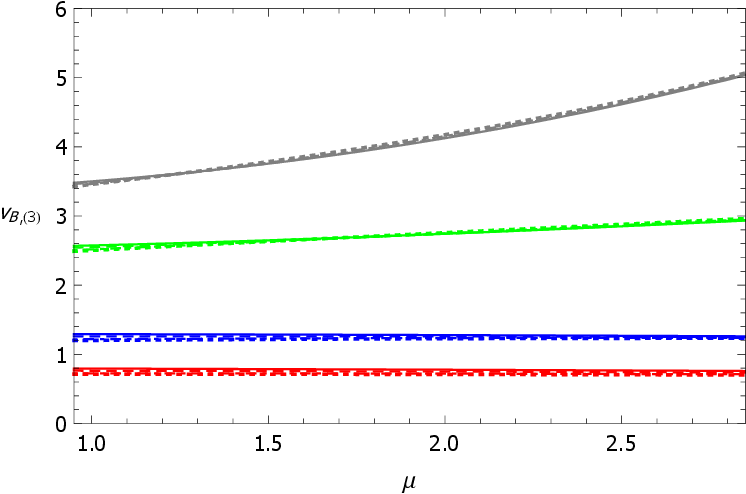}}
\subfigure[$v_{B,(1)}$ vs. $\mu$ for $T=2.1$]{
\label{figvb5d}
\includegraphics[width=0.3\linewidth]{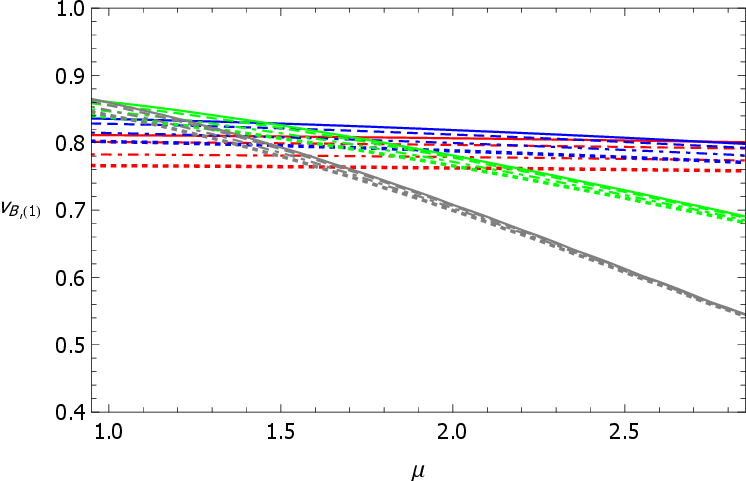}}
\subfigure[$v_{B,(2)}$ vs. $\mu$ for $T=2.1$]{
\label{figvb5e}
\includegraphics[width=0.3\linewidth]{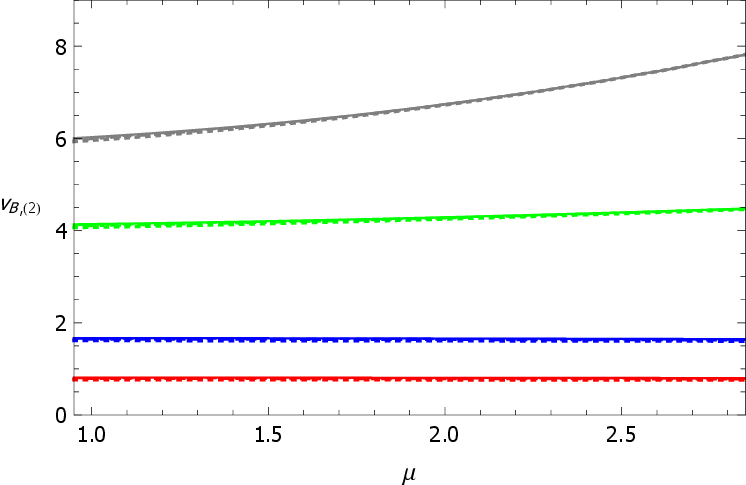}}
\subfigure[$v_{B,(3)}$ vs. $\mu$ for $T=2.1$]{
\label{figvb5f}
\includegraphics[width=0.3\linewidth]{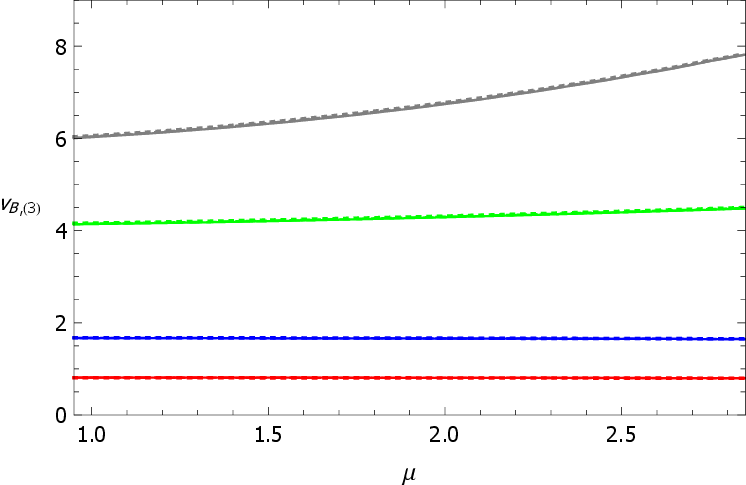}}
\caption{Butterfly velocities $v_{B,(i)}$ as functions of $\mu$. The first and second rows correspond to $T=1.05$ and $2.1$, respectively. The black, red, blue, and gray curves denote $\nu=1,1.5,3$, and $4.5$, while the solid, dashed, dot-dashed, and dotted curves correspond to $c_B=0,-1,-3$, and $-5$. We use $R_{gg}=1.16\,\text{GeV}^2$, $p=0.273\,\text{GeV}^4$, and $q_3=5\,\text{GeV}^2$, with $[\mu]=[T]=[c_B]^{1/2}=\text{GeV}$.}
\label{figvb5}
\end{figure*}

Figures~\ref{figvb4} and~\ref{figvb5} show the temperature and chemical potential dependence of the directional butterfly velocities. As seen in Figure~\ref{figvb4}, all three components increase with temperature, as in the isotropic case.
Figure~\ref{figvb5} shows that, once anisotropy is introduced, the transverse components $v_{B,(2)}$ and $v_{B,(3)}$ change from decreasing functions of the chemical potential to monotonically increasing ones. A larger $\nu$ further strengthens the positive dependence of the transverse velocities on both temperature and chemical potential, indicating that the anisotropy-induced enhancement dominates over the suppressing effects of the other parameters.
The chemical-potential dependence of $v_{B,(1)}$, however, is controlled jointly by the anisotropy and magnetic field. For weak anisotropy and a strong magnetic field, $v_{B,(1)}$ increases with $\mu$, as can also be seen by comparing Figures~\ref{figvb3a} and~\ref{figvb3d}; for strong anisotropy, it decreases with $\mu$.

The main results for the butterfly velocities in the doubly anisotropic holographic system can be summarized as follows:

\begin{itemize}
\item In an isotropic QGP, the butterfly velocity increases with temperature and approaches the conformal benchmark. It decreases with chemical potential~\cite{Lilani:2025wnd}.

\item In the doubly anisotropic QGP, the primary anisotropy is the dominant source of directional splitting in the butterfly velocities, producing a substantial difference between the longitudinal and transverse components. The magnetic field produces a much smaller splitting, as reflected by the small difference between the two transverse components.

\item In the doubly anisotropic QGP with magnetic catalysis, the two transverse velocities increase monotonically with $\nu$, whereas the dependence of the longitudinal component on $\nu$ is modulated by the chemical potential. The magnetic field suppresses the longitudinal butterfly velocity but has only a weak effect on the transverse components.

\item Two additional features are noteworthy. First, the butterfly velocity along the anisotropic direction can exceed the isotropic conformal benchmark. As discussed in Ref.~\cite{Gursoy:2020kjd}, this may arise because the present metric lies outside the conformal class considered in Refs.~\cite{Giataganas:2017koz,Jahnke:2017iwi}. Second, in the coordinate normalization used here, both transverse components may exceed unity. These values should not be identified directly with a violation of relativistic causality without first imposing a common physical normalization of the anisotropic boundary coordinates~\cite{Balasubramanian:2019stt, Basu:2025exh, Basu:2026wzp}.

\item The monotonic increase of the butterfly velocities with temperature is not altered by the anisotropy or magnetic field. Their response to the chemical potential, however, is direction dependent: unlike the isotropic butterfly velocity, which decreases with $\mu$, the two transverse components increase monotonically, while the longitudinal response is jointly controlled by the anisotropy and magnetic field.
\end{itemize}

\section{Drag Force}
\label{sec:LOSS}
In this section, we study the drag force in the background metric of the doubly anisotropic magnetized QGP. A heavy quark moving at constant velocity through a strongly coupled plasma experiences a drag force and loses energy via interactions with the medium. The holographic calculation of the drag force in isotropic $\mathcal{N}=4$ supersymmetric Yang--Mills (SYM) theory was pioneered in Refs.~\cite{Gubser:2006bz,Herzog:2006gh}. The corresponding Langevin equation can be written as
\begin{eqnarray}
\frac{d p}{d t} = - \mu_D p + f.
\end{eqnarray}
where $p$ is the momentum of the heavy quark, $\mu_D$ is the drag coefficient, and $f$ is an external force. In the steady-state limit, $d p/d t=0$, the external force balances the drag force.

Unlike in an isotropic plasma, the drag coefficient in an anisotropic medium is generally not a scalar; the force experienced by a heavy-quark probe depends on its direction of motion. In the present doubly anisotropic background, $f_{x_1}$, $f_{x_2}$, $f_{x_3}$ are generally unequal.

Now, we provide a short derivation of the Holographic heavy-quark drag formula in  the a doubly anisotropic EMD model. Note that the drag force is typically computed analytically in the string frame by evaluating the Nambu-Goto action. We therefore transform the Einstein frame metric, Eq.~(\ref{metricansatz}), into the string frame by a Weyl transformation, where the metric can be simplified to
\begin{eqnarray}
d s_{(S)}^2 &=& \tilde{G}_{t\, t} dt^2 + \sum_{i =1}^{3} \tilde{G}_{x_i\, x_i} d x_i^2 + \tilde{G}_{z\, z}d z^2.
\label{metric_ansatz 2}
\end{eqnarray}
where $\tilde{G}_{t t}=-z^{-2}\mathfrak{b}_s(z)g(z)$, $\tilde{G}_{x_1 x_1}=z^{-2}\mathfrak{b}_s(z)$, $\tilde{G}_{x_2 x_2}=z^{-2/\nu}\mathfrak{b}_s(z)$, $\tilde{G}_{x_3 x_3}=z^{-2/\nu}e^{c_Bz^2}\mathfrak{b}_s(z)$, and $\tilde{G}_{z z}=z^{-2}\mathfrak{b}_s(z)/g(z)$. $\mathfrak{b}_s(z) = \mathfrak{b}(z) e^{ \sqrt{2/3} \phi(z)}$ denotes the warp factor in the string frame. The above expression carries the same bulk geometric information as Eq.~(\ref{Einstein metric}) in the previous section~\ref{sec:EMD}, differing only in the choice between the string and Einstein frames.

We now consider an external heavy quark moving through the strongly coupled plasma along the $x_i$ direction with constant velocity $v$:
\begin{eqnarray}
\label{moving}
t= \tau,\;\;z= \sigma,\;\; x_i = v \tau + \xi_{x_i}(z),\;\; x_{j\neq i}=0.
\end{eqnarray}
Here, $x_i\in\{x_1,x_2,x_3\}$ denotes the direction of motion, while the worldsheet is localized in the remaining spatial directions. The function $\xi_{x_i}(z)$ is chosen to vanish at the boundary $z=0$ so that the quark moves there with constant velocity $v$.

The Nambu--Goto action in the string frame is
\begin{eqnarray}
S&=& -\frac{1}{2\pi\alpha'} \int d\tau d\sigma \mathcal{L}\\
 &=& -\frac{1}{2\pi\alpha'} \int d\tau d\sigma \sqrt{-\det g_{\alpha\beta}},\;\;\;\; g_{\alpha\beta} = \tilde{G}_{\mu\nu}\frac{\partial X^{\mu}}{\partial\sigma^{\alpha}}\frac{\partial X^{\nu}}{\partial\sigma^{\beta}} \nonumber.
\end{eqnarray}
where $g_{\alpha\beta}$ is the induced world-sheet metric, while $\tilde{G}_{\mu\nu}$ and $X^\mu$ denote the target-space metric and embedding coordinates, respectively.

Substituting the embedding~(\ref{moving}) into the metric~(\ref{metric_ansatz 2}) gives the induced world-sheet metric components
\begin{eqnarray}
g_{t t} &=& \tilde{G}_{t t} + \tilde{G}_{x_i x_i}v^2,\nonumber\\
g_{t z}=g_{z t} &=& \tilde{G}_{x_i x_i}v\,\xi_{x_i}',\nonumber\\
g_{z z} &=& \tilde{G}_{x_i x_i}\xi_{x_i}'^2 + \tilde{G}_{z z}.
\end{eqnarray}
where $\xi_{x_i}'=d \xi_{x_i}/d \sigma$.

The corresponding Lagrangian density is
\begin{eqnarray}
\mathcal{L}=\sqrt{ -\tilde{G}_{t t} \tilde{G}_{z z}-\tilde{G}_{z z}\tilde{G}_{x_i x_i }v^2 - \tilde{G}_{t t}\tilde{G}_{x_i x_i}\xi_{x_i}'^2}.
\end{eqnarray}
Because $\mathcal{L}$ does not depend explicitly on $\xi_{x_i}$, its conjugate momentum is conserved:
\begin{eqnarray}
\Pi_{x_i}
&=& -\frac{\tilde{G}_{t t}\tilde{G}_{x_i x_i}}{\sqrt{-\det g_{\alpha\beta}}}\,\xi_{x_i}'\\
&=& \frac{-\tilde{G}_{t t}\tilde{G}_{x_i x_i}\xi_{x_i}'}{\sqrt{-\tilde{G}_{t t}\tilde{G}_{z z}-\tilde{G}_{z z}\tilde{G}_{x_i x_i}v^2-\tilde{G}_{t t}\tilde{G}_{x_i x_i}(\xi_{x_i}')^2}},\;\;\;\nonumber
\end{eqnarray}
which leads to
\begin{eqnarray}
\xi_{x_i}' = \sqrt{\frac{-\tilde{G}_{z z}(\tilde{G}_{tt} + \tilde{G}_{x_i x_i}v^2)\Pi_{x_i}^2}{\tilde{G}_{t t}\tilde{G}_{x_i x_i}(\tilde{G}_{t t}\tilde{G}_{x_i x_i}+ \Pi_{x_i}^2 )}}.
\end{eqnarray}
The profile $\xi_{x_i}'$ must remain real along the entire string. Consequently, the numerator and denominator of the fraction under the square root must have the same sign for all $z$. The numerator vanishes at $z=z_c$, where
\begin{eqnarray}
\tilde{G}_{z z}(\tilde{G}_{tt} + \tilde{G}_{x_i x_i} v^2)|_{z= z_c}=0.
\label{sol zc}
\end{eqnarray}
In practice, the relevant solution follows from the expression in parentheses. The coordinate $z_c$ is not merely an intermediate variable; it identifies the characteristic radial depth associated with energy dissipation along the trailing string. Requiring regularity at $z_c$ also fixes the conserved momentum as
\begin{eqnarray}
\Pi_{x_i} = \sqrt{ - \tilde{G}_{t t}\tilde{G}_{x_i x_i}}|_{z= z_c}.
\end{eqnarray}

The radial flux of momentum in the $x_i$ direction can be written as

\begin{eqnarray}
P_{x_i} = -\frac{1}{2 \pi \alpha'} \tilde{G}_{{x_i} \nu}\partial^\beta X^\nu =-\frac{1}{2 \pi \alpha'}\frac{\tilde{G}_{t t}\tilde{G}_{{x_i} {x_i} } }{-\det g_{\alpha\beta}}\xi_{x_i}',
\end{eqnarray}
and is related to the momentum loss of the heavy quark through
\begin{eqnarray}
\frac{d p_{x_i}}{d t }=\sqrt{-\det g_{\alpha\beta}}P_{x_i} = -\frac{1}{2 \pi \alpha'}\Pi_{x_i} ,
\end{eqnarray}
Choosing the physical trailing-string solution, for which momentum flows from the boundary toward the horizon, gives the drag force along $x_i$:
\begin{eqnarray}
f_{x_i}=\frac{d p_{x_i}}{d t }= -\frac{1}{2 \pi \alpha'}\tilde{G}_{x_i x_i} v|_{z= z_c} .
\label{dragall}
\end{eqnarray}
Thus, the drag force in an arbitrary principal direction is obtained by first solving Eq.~(\ref{sol zc}) for $z_c$ and then substituting the result into Eq.~(\ref{dragall}).

Substituting the metric components from Eq.~(\ref{metric_ansatz 2}) into Eq.~(\ref{dragall}) yields the directional drag forces acting on a heavy-quark probe in the doubly anisotropic QGP:
\begin{eqnarray}
f_{x_1}&=&-\frac{\sqrt{\lambda}}{2 \pi} {z_c^{-2}}\mathfrak{b}_s(z_c)v \Big|_{ g(z_c)=v^2 },
\label{drag x1}\\
f_{x_2}&=&-\frac{\sqrt{\lambda}}{2 \pi}z_c^{-\frac{2}{\nu}}\mathfrak{b}_s(z_c)v
\Big|_{(z_c)^{{2}/{\nu}-2}g(z_c)=v^2 },
\label{drag x2}\\
f_{x_3}&=&-\frac{\sqrt{\lambda}}{2 \pi}{z_c}^{-\frac{2}{\nu}} e^{c_B z_c^2}\mathfrak{b}_s(z_c)v
\Big|_{{z_c}^{{2}/{\nu}-2}e^{-c_B z_c^2}g(z_c)=v^2}.
\label{drag x3}
\end{eqnarray}
Where, for a fixed quark velocity \(v\), \(z_c\) is determined for each direction by Eq.~(\ref{sol zc}), with \(\tilde G_{x_i x_i}\) taken along that direction. \(\sqrt{\lambda}\equiv L^2/\alpha'=1/\alpha'\) is dimensionless. The drag force is determined entirely by the metric components evaluated at \(z_c\), highlighting the geometric origin of energy loss in a strongly coupled plasma.

Although these expressions are compact, the corresponding values of $z_c$ cannot, in general, be obtained analytically. We therefore determine $z_c$ and the drag forces numerically.

Before presenting the numerical results, we recall the corresponding expressions for an isotropic medium and for $\mathcal{N}=4$ SYM theory. Setting $c_B=0$ and $\nu=1$ gives the drag force for a heavy-quark probe in an isotropic QGP:
\begin{eqnarray}
\label{drag22}
f_{{iso}}&=&-\frac{\sqrt{\lambda}}{2 \pi} {z_c^{-2}}\mathfrak{b}_s(z_c)v \Big|_{v^2 =g_{{iso}}(z_c)},
\end{eqnarray}
where $g_{{iso}}(z_c)$ denotes the isotropic blackening function evaluated at $z_c$ in the absence of magnetic catalysis. Throughout the following discussion, $f_{{iso}}$ denotes this zero-field isotropic reference, defined by $\nu=1$ and $c_B=0$. The $\mathcal{N}=4$ SYM result is~\cite{Gubser:2006bz,Herzog:2006gh}
\begin{eqnarray}
f_{\text{SYM}}&=&-\frac{\pi\sqrt{\lambda}}{2} T^2 \frac{v}{\sqrt{1-v^2}}.
\label{drag sym}
\end{eqnarray}

\begin{figure*}[t]
\centering
\subfigure[$f_{{iso}}/f_{\text{SYM}}$ vs. $v$]{
\label{figdrag1a}
\includegraphics[width=0.3\linewidth]{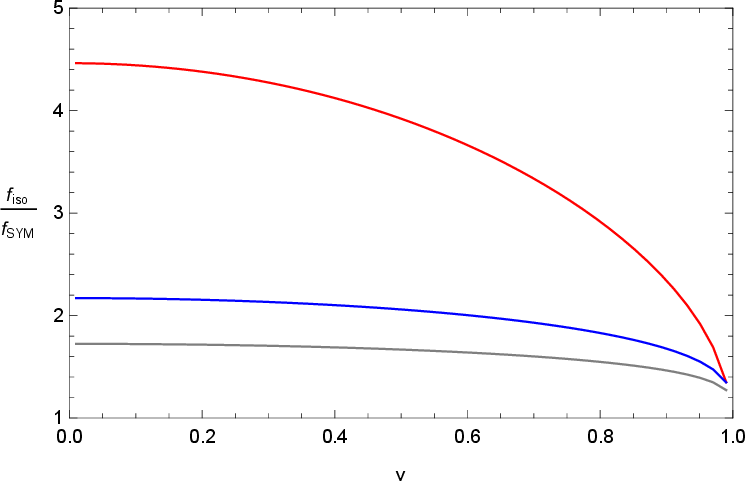}}
\subfigure[$f_{{iso}}/f_{\text{SYM}}$ vs. $T$]{
\label{figdrag1b}
\includegraphics[width=0.3\linewidth]{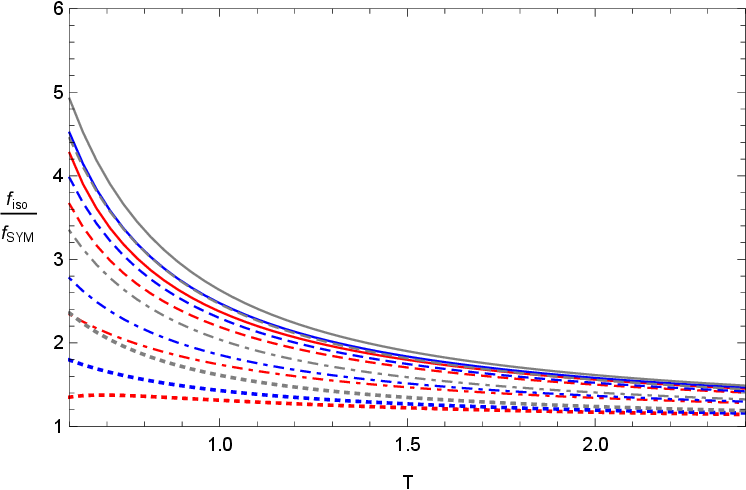}}
\subfigure[$f_{{iso}}/f_{\text{SYM}}$ vs. $\mu$]{
\label{figdrag1c}
\includegraphics[width=0.3\linewidth]{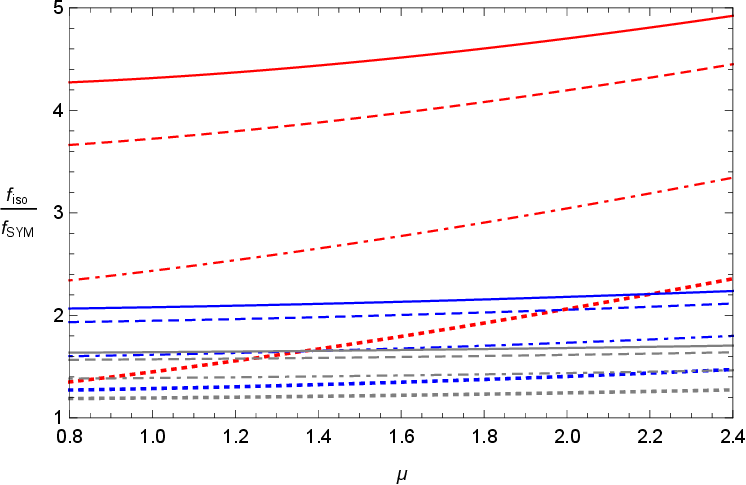}}
\caption{The ratio $f_{{iso}}/f_{\text{SYM}}$ as a function of (a) the probe velocity $v$, (b) the temperature $T$, and (c) the chemical potential $\mu$. In panel (a), the red, blue, and gray curves correspond to $(\mu,T)=(0.8,0.6)$, $(1.6,1.2)$, and $(2.4,1.8)$, respectively. In panel (b), the red, blue, and gray curves correspond to $\mu=0.8,1.6,2.4$; in panel (c), they correspond to $T=0.6,1.2,1.8$; the solid, dashed, dot-dashed, and dotted curves denote $v=0.3,0.6,0.9$, and $0.99$. We use $R_{gg}=1.16\,\text{GeV}^2$, $p=0.273\,\text{GeV}^4$, and $q_3=5\,\text{GeV}^2$, with $[\mu]=[T]=\text{GeV}$.}
\label{figdrag1}
\end{figure*}

We first compare the drag force in an isotropic QGP ($\nu=1$) at zero magnetic field ($c_B=0$) with the corresponding SYM result. As shown in
Figure~\ref{figdrag1},  the ratio $f_{{iso}}/f_{\text{SYM}}$ is always greater than unity,  reflecting the enhancing effect of the warp factor arising from non-conformal effects in the holographic model and thus indicating that the QGP drag is generally stronger. Figure~\ref{figdrag1a} shows a monotonic decrease of the ratio with increasing $v$; this decrease becomes less pronounced at high temperatures and chemical potentials. Figure~\ref{figdrag1b} demonstrates that, for the different fixed chemical potentials considered, the drag force ratio decreases with increasing temperature, approaching unity in the high-temperature limit. This indicates that the increasing trend of the drag force with temperature is slower in the QGP than in the SYM system. In contrast, as shown in Figure~\ref{figdrag1c}, the drag force ratio  increases overall with increasing chemical potential $\mu$. 

\begin{figure*}[t]
\centering
\subfigure[$f_{{x}_1}/f_{{iso}}$ vs. $\nu$ for $(\mu,\; T) \;=\;(0.95,\;1.05)$]{
\label{figdrag2a}
\includegraphics[width=0.3\linewidth]{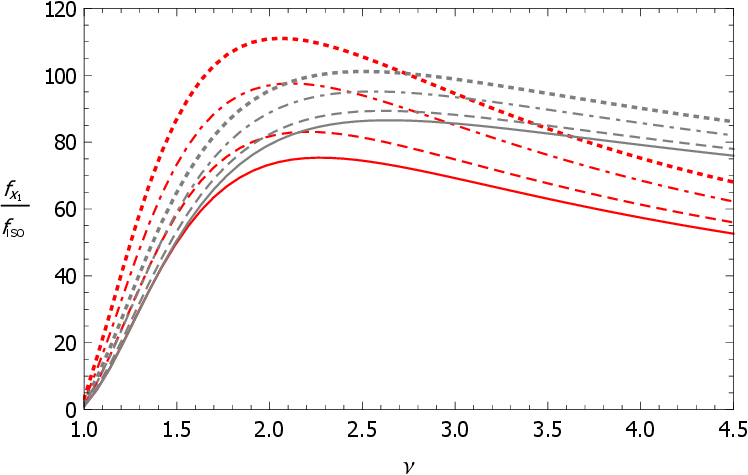}}
\subfigure[$f_{{x}_2}/f_{{iso}}$ vs. $\nu$ for $(\mu,\; T) \;=\;(0.95,\;1.05)$]{
\label{figdrag2b}
\includegraphics[width=0.3\linewidth]{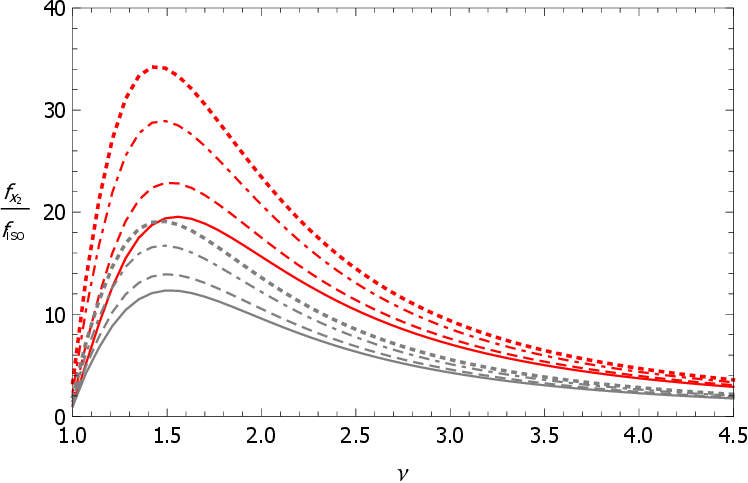}}
\subfigure[$f_{{x}_3}/f_{{iso}}$ vs. $\nu$ for $(\mu,\; T)\;=\;(0.95,\;1.05)$]{
\label{figdrag2c}
\includegraphics[width=0.3\linewidth]{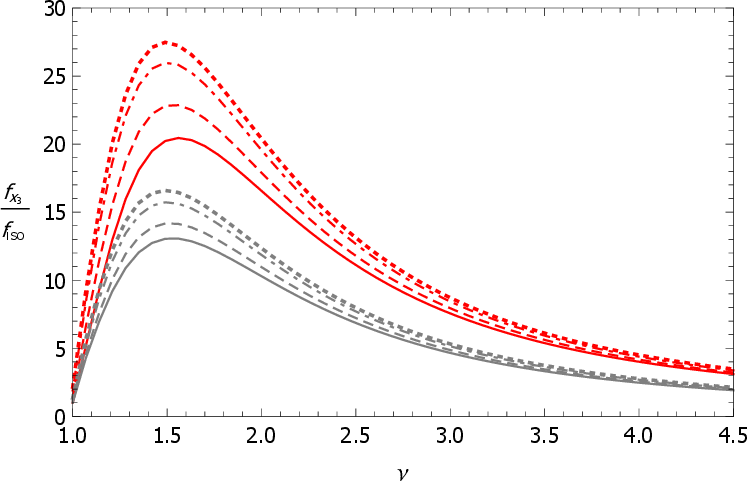}}
\subfigure[$f_{{x}_1}/f_{{iso}}$ vs. $\nu$ for $(\mu, T) \;=\;(1.9, 2.1)$]{
\label{figdrag2d}
\includegraphics[width=0.3\linewidth]{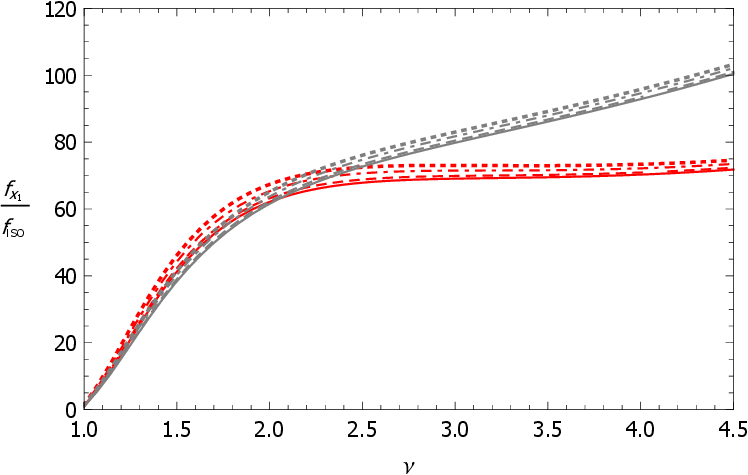}}
\subfigure[$f_{{x}_2}/f_{{iso}}$ vs. $\nu$ for $(\mu,\; T)\;=\;(1.9,\;2.1)$]{
\label{figdrag2e}
\includegraphics[width=0.3\linewidth]{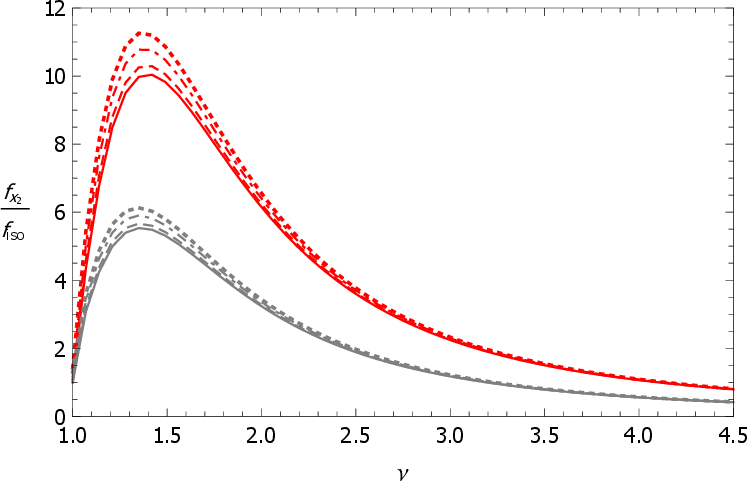}}
\subfigure[$f_{{x}_3}/f_{{iso}}$ vs. $\nu$ for $(\mu,\; T) \;=\;(1.9,\;2.1)$]{
\label{figdrag2f}
\includegraphics[width=0.3\linewidth]{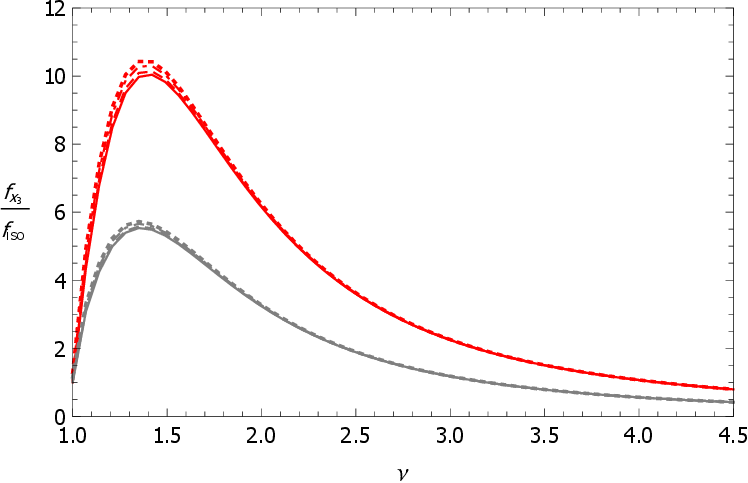}}
\caption{Ratios $f_{x_i}/f_{{iso}}$ as functions of the primary anisotropy parameter $\nu$. The first and second rows correspond to $(\mu,T)=(0.95,1.05)$ and $(1.9,2.1)$, respectively. The red and gray curves denote $v=0.3$ and $0.9$, while the solid, dashed, dot-dashed, and dotted curves correspond to $c_B=0,-1,-3$, and $-5$. We use $R_{gg}=1.16\,\text{GeV}^2$, $p=0.273\,\text{GeV}^4$, and $q_3=5\,\text{GeV}^2$, with $[\mu]=[T]=[c_B]^{1/2}=\text{GeV}$.}
\label{figdrag2}
\end{figure*}

\begin{figure*}[t]
\centering
\subfigure[$f_{{x}_1}/f_{{iso}}$ vs. $|c_B|$ for $(\mu,\; T) \;=\;(0.95,\;1.05)$]{
\label{figdrag3a}
\includegraphics[width=0.3\linewidth]{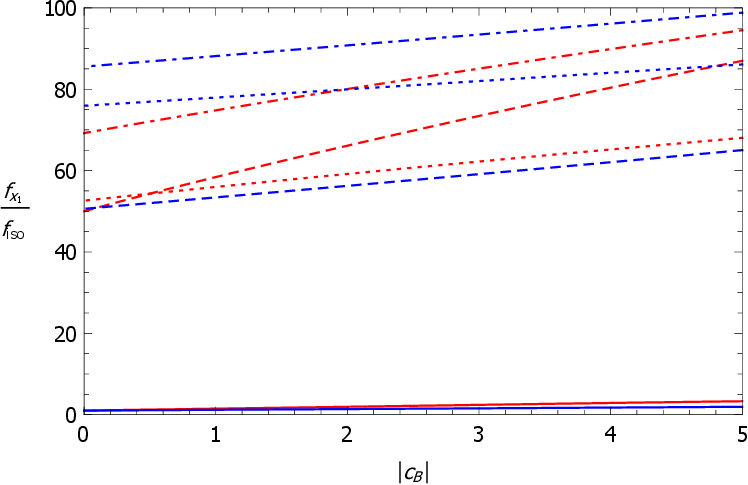}}
\subfigure[$f_{{x}_2}/f_{{iso}}$ vs. $|c_B|$ for $(\mu,\; T) \;=\;(0.95,\;1.05)$]{
\label{figdrag3b}
\includegraphics[width=0.3\linewidth]{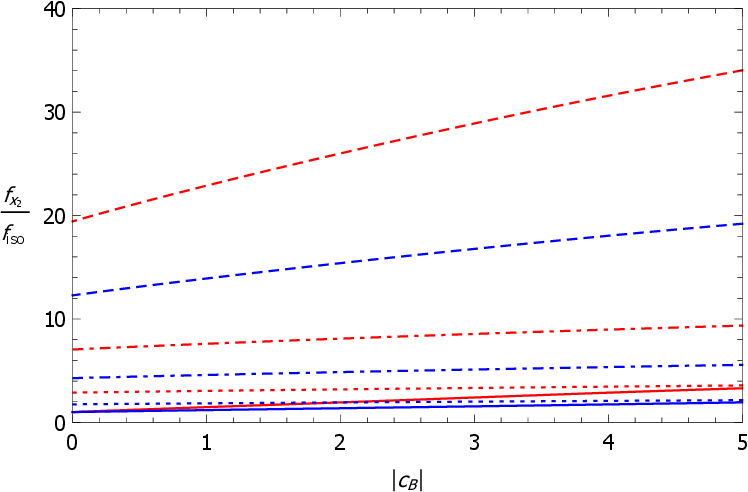}}
\subfigure[$f_{{x}_3}/f_{{iso}}$ vs. $|c_B|$ for $(\mu,\; T) \;=\;(0.95,\;1.05)$]{
\label{figdrag3c}
\includegraphics[width=0.3\linewidth]{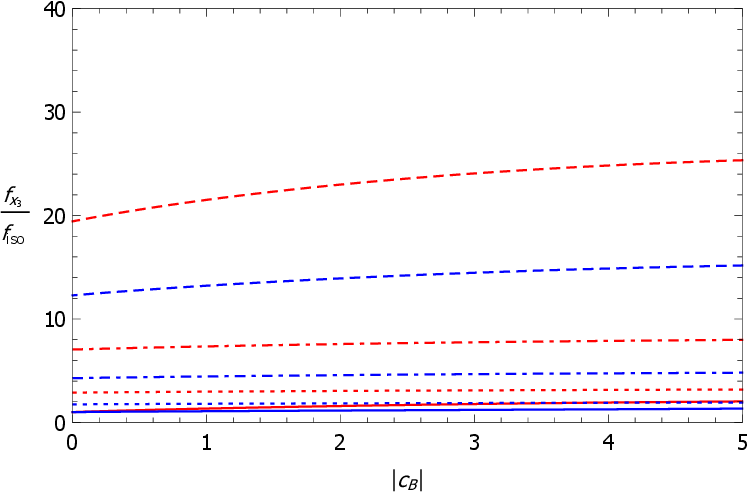}}
\subfigure[$f_{{x}_1}/f_{{iso}}$ vs. $|c_B|$ for $(\mu,\; T) \;=\;(1.9,\;2.1)$]{
\label{figdrag3d}
\includegraphics[width=0.3\linewidth]{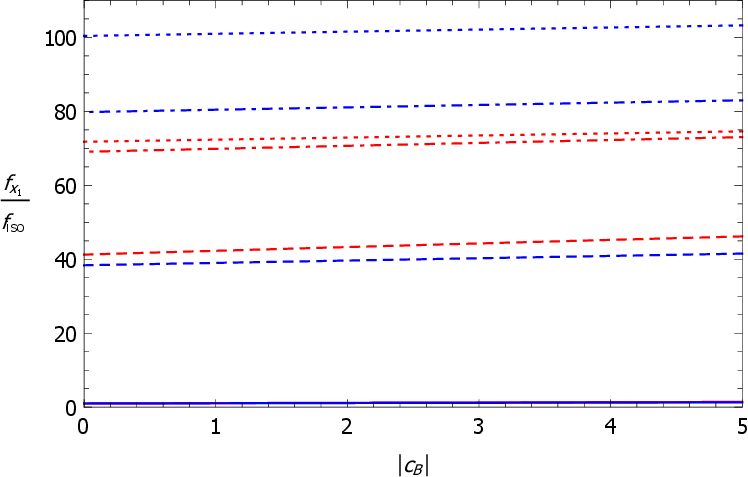}}
\subfigure[$f_{{x}_2}/f_{{iso}}$ vs. $|c_B|$ for $(\mu,\; T) \;=\;(1.9,\;2.1)$]{
\label{figdrag3e}
\includegraphics[width=0.3\linewidth]{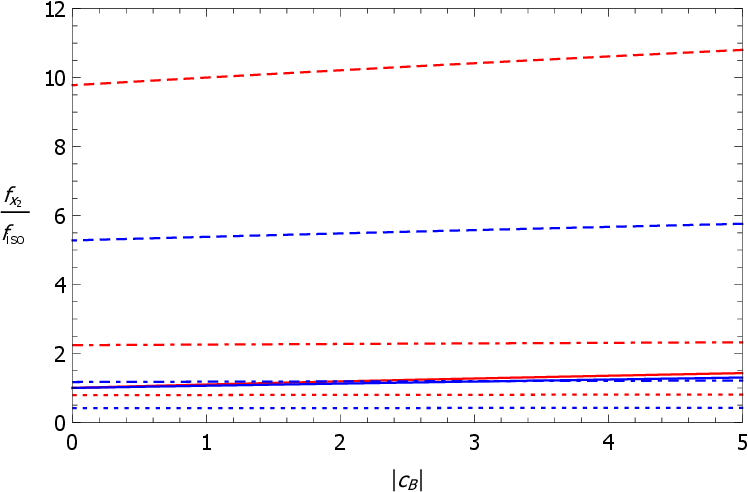}}
\subfigure[$f_{{x}_3}/f_{{iso}}$ vs. $|c_B|$ for $(\mu,\; T) \;=\;(1.9,\;2.1)$]{
\label{figdrag3f}
\includegraphics[width=0.3\linewidth]{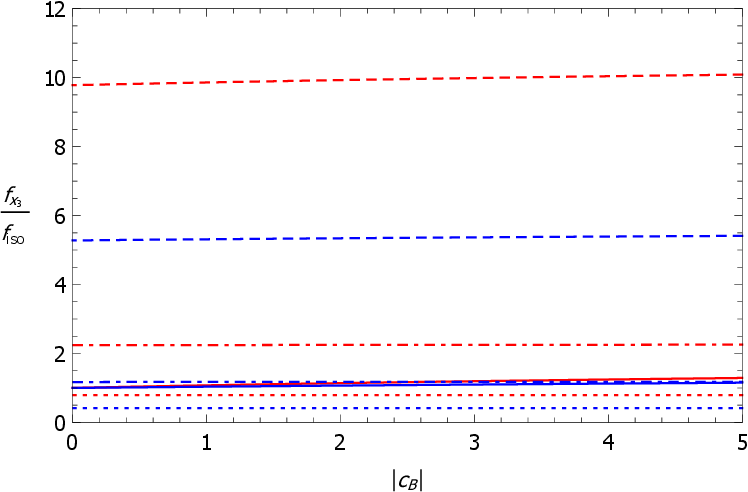}}
\caption{Ratios $f_{x_i}/f_{{iso}}$ as functions of $|c_B|$. The first and second rows correspond to $(\mu,T)=(0.95,1.05)$ and $(1.9,2.1)$, respectively. The red and blue curves denote  $v=0.3$ and $0.9$, while the solid, dashed, dot-dashed and dotted curves correspond to $\nu=1$, $1.5$, $3$ and $4.5$.  We use $R_{gg}=1.16\,\text{GeV}^2$, $p=0.273\,\text{GeV}^4$, and $q_3=5\,\text{GeV}^2$, with $[\mu]=[T]=[c_B]^{1/2}=\text{GeV}$.}
\label{figdrag3}
\end{figure*}

Figure~\ref{figdrag2} shows the ratios $f_{x_i}/f_{{iso}}$ as functions of the anisotropy parameter $\nu$ for several magnetic fields and quark velocities at fixed temperature and chemical potential. Overall, the drag force in the doubly anisotropic QGP exhibits pronounced anisotropy, and the anisotropy parameter plays a more dominant role than the magnetic field. Specifically, the drag force along the direction parallel to the anisotropy is significantly stronger than that along the two transverse directions, while the difference between the two transverse directions is relatively small. From Figures~\ref{figdrag2a}--\ref{figdrag2c} in the first row, it can be seen that at lower temperature and chemical potential, for quarks moving along any one of the three directions in the doubly anisotropic QGP, the drag force ratio increases rapidly as $\nu$ increases, reaches a peak around $\nu \approx 1.5$--$2.0$, and then decreases slowly, but always remains greater than $1$. This indicates that in the QGP near the phase-transition critical point, moderate anisotropy most significantly enhances the drag forces, whereas for stronger anisotropy the enhancement effect is instead weakened. However, in the doubly anisotropic QGP at high temperature and chemical potential, as seen in Figures~\ref{figdrag2d}--\ref{figdrag2f} in the second row, the dependence of the drag force on $\nu$ exhibits different features: first, $f_{x_1}$ increases monotonically with $\nu$; Second, \(f_{x_2}\) and \(f_{x_3}\) also first increase and then decrease, reaching a peak at \(\nu\approx 1.5\), and eventually fall below the isotropic reference value for stronger anisotropy. This means that in the extreme environment of high temperature and chemical potential, strong anisotropy enhances the drag force parallel to the anisotropy direction, but suppresses the drag force transverse to the anisotropy direction. In addition, for the anisotropy parameters and quark velocities considered at fixed temperature and chemical potential, increasing the external magnetic field strength \(|c_B|\) enhances the drag-force ratios in all three directions in the doubly anisotropic QGP, as corroborated in Figure~\ref{figdrag3}. Furthermore, a comparison of the numerical results for $f_{x_2}$ and $f_{x_3}$ reveals that, with other parameters held fixed, the presence of an external magnetic field leads to a drag force parallel to the external magnetic field direction, \(f_{x_3}\), that is smaller than that perpendicular to it, \(f_{x_2}\).

\begin{figure*}[t]
\centering
\subfigure[$f_{{x}_1}/{\sqrt{\lambda}}$ vs. $v$ for $(\mu,\; T)\;=\;(0.95,\;1.05)$]{
\label{figdrag4a}
\includegraphics[width=0.3\linewidth]{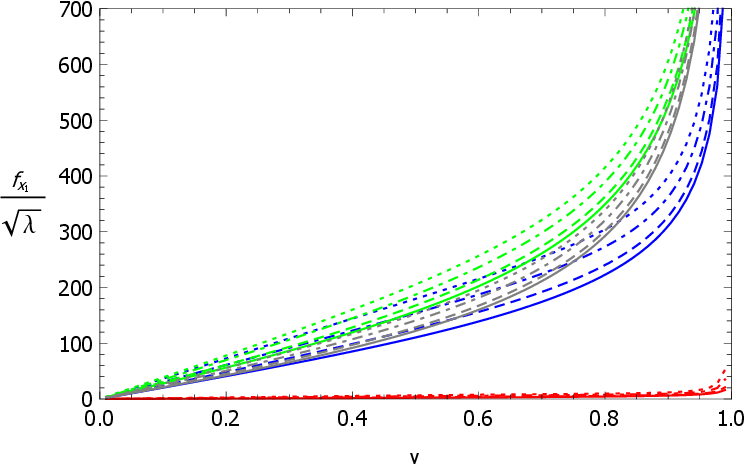}}
\subfigure[$f_{{x}_2}/{\sqrt{\lambda}}$ vs. $v$ for $(\mu,\; T)\;=\;(0.95,\;1.05)$]{
\label{figdrag4b}
\includegraphics[width=0.3\linewidth]{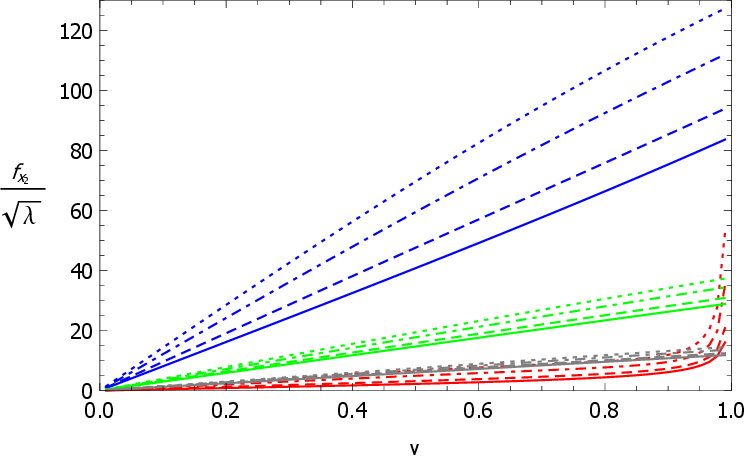}}
\subfigure[$f_{{x}_3}/{\sqrt{\lambda}}$ vs. $v$ for $(\mu,\; T)\;=\;(0.95,\;1.05)$]{
\label{figdrag4c}
\includegraphics[width=0.3\linewidth]{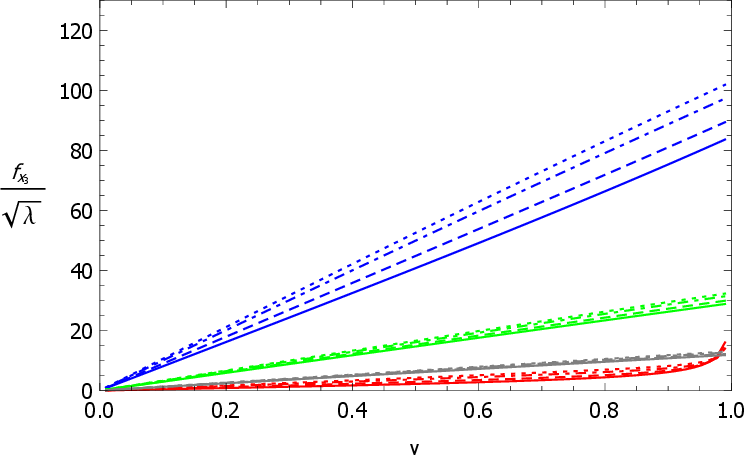}}
\subfigure[$f_{{x}_1}/{\sqrt{\lambda}}$ vs. $v$ for $(\mu,\; T)\;=\;(1.9,\; 2.1)$]{
\label{figdrag4d}
\includegraphics[width=0.3\linewidth]{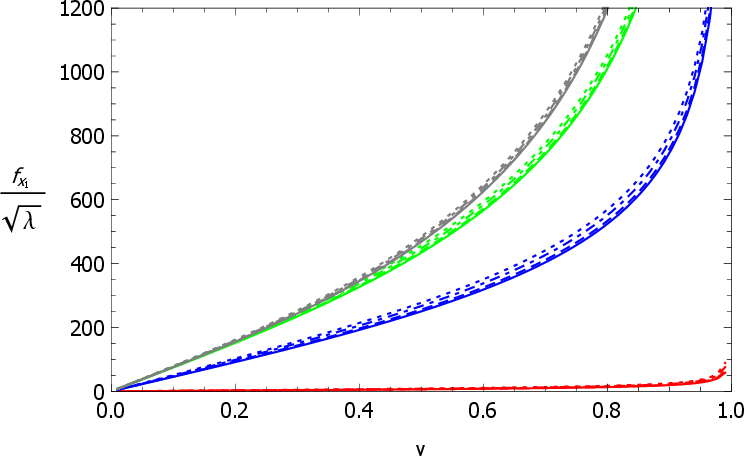}}
\subfigure[$f_{{x}_2}/{\sqrt{\lambda}}$ vs. $v$ for $(\mu,\; T)\;=\;(1.9,\;2.1)$]{
\label{figdrag4e}
\includegraphics[width=0.3\linewidth]{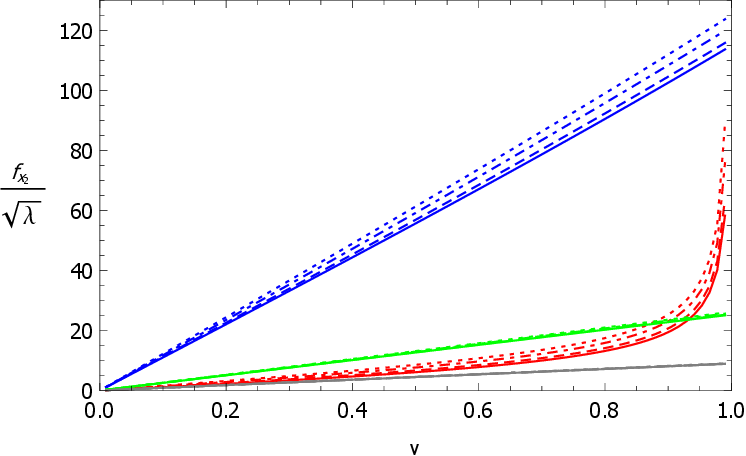}}
\subfigure[$f_{{x}_3}/{\sqrt{\lambda}}$ vs. $v$ for $(\mu,\; T)\;=\;(1.9,\;2.1)$]{
\label{figdrag4f}
\includegraphics[width=0.3\linewidth]{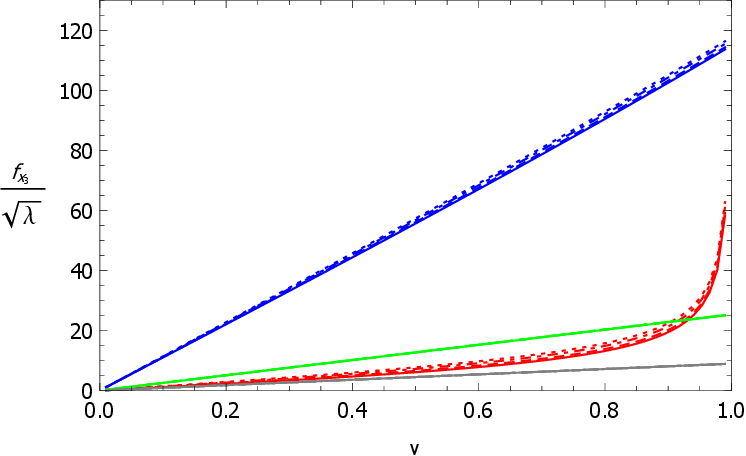}}
\caption{Directional drag forces $f_{x_i}/\sqrt{\lambda}$ as functions of the quark velocity $v$. The first and second rows correspond to $(\mu,T)=(0.95,1.05)$ and $(1.9,2.1)$, respectively. The red, blue, and gray curves denote $\nu=1$, $1.5$, and $3$, while the solid, dashed, dot-dashed, and dotted curves correspond to $c_B=0,-1,-3$, and $-5$. We use $R_{gg}=1.16\,\text{GeV}^2$, $p=0.273\,\text{GeV}^4$, and $q_3=5\,\text{GeV}^2$, with $[\mu]=[T]=[c_B]^{1/2}=[f_{x_i}]^{1/2}=\text{GeV}$.}
\label{figdrag4}
\end{figure*}

\begin{figure*}[t]
\centering
\subfigure[$f_{{x}_1}/{\sqrt{\lambda}}$ vs. $T$ for $v=0.3$]{
\label{figdrag5a}
\includegraphics[width=0.3\linewidth]{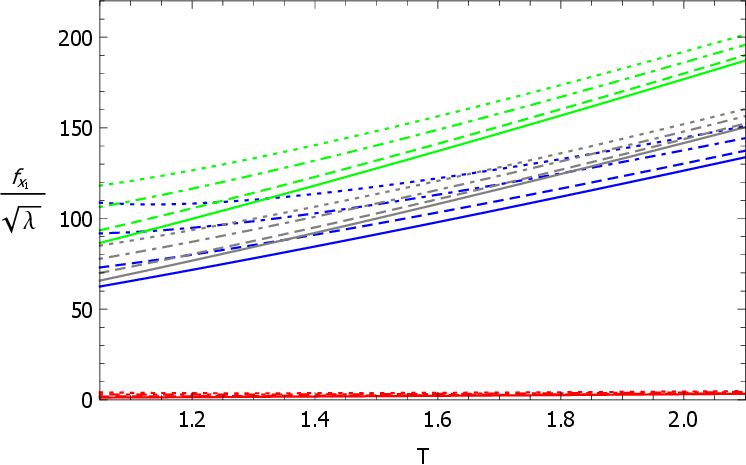}}
\subfigure[$f_{{x}_2}/{\sqrt{\lambda}}$ vs. $T$ for $v=0.3$]{
\label{figdrag5b}
\includegraphics[width=0.3\linewidth]{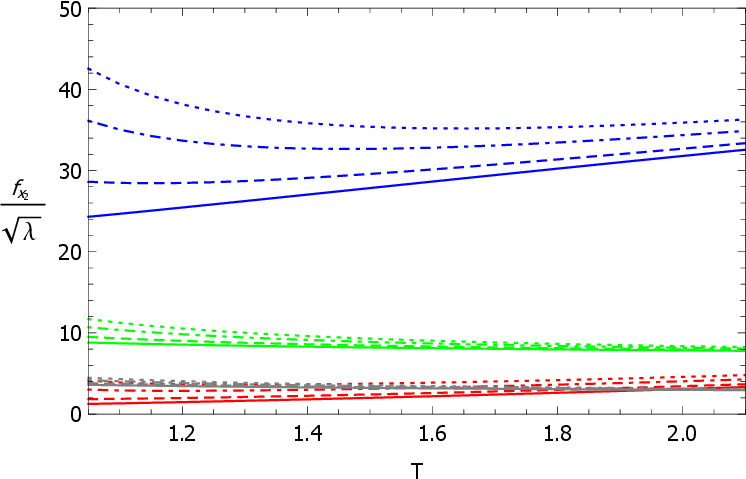}}
\subfigure[$f_{{x}_3}/{\sqrt{\lambda}}$ vs. $T$ for $v=0.3$]{
\label{figdrag5c}
\includegraphics[width=0.3\linewidth]{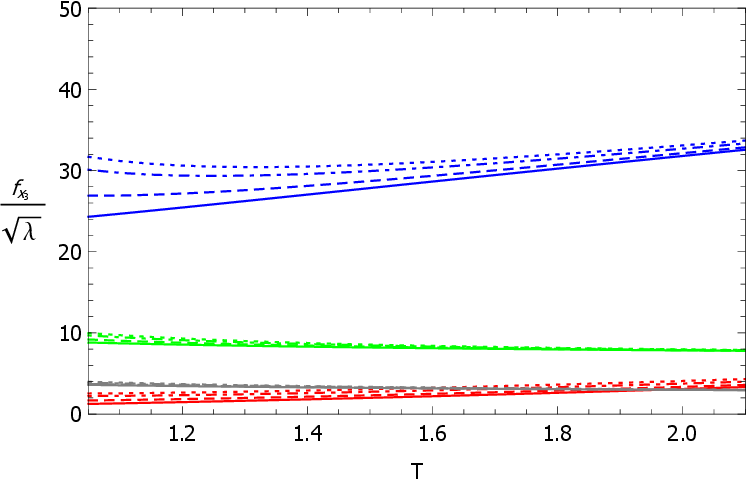}}
\subfigure[$f_{{x}_1}/{\sqrt{\lambda}}$ vs. $T$ for $v=0.9$]{
\label{figdrag5d}
\includegraphics[width=0.3\linewidth]{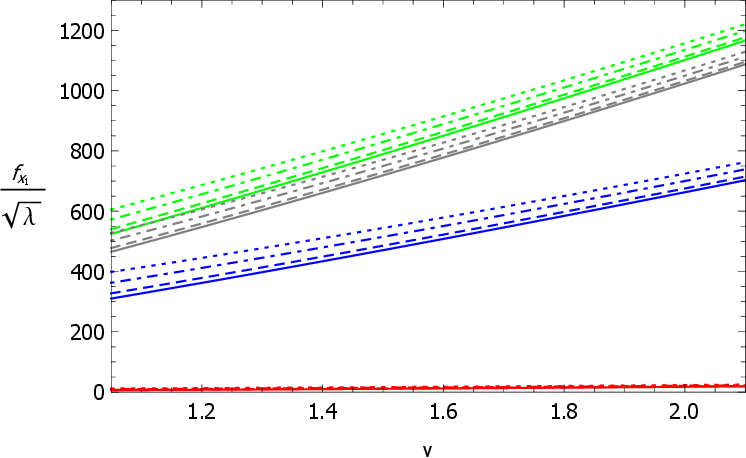}}
\subfigure[$f_{{x}_2}/{\sqrt{\lambda}}$ vs. $T$ for $v=0.9$]{
\label{figdrag5e}
\includegraphics[width=0.3\linewidth]{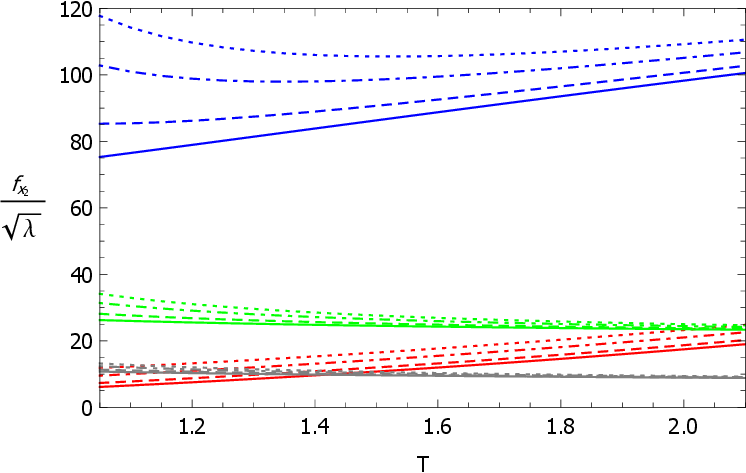}}
\subfigure[$f_{{x}_3}/{\sqrt{\lambda}}$ vs. $T$ for $v=0.9$]{
\label{figdrag5f}
\includegraphics[width=0.3\linewidth]{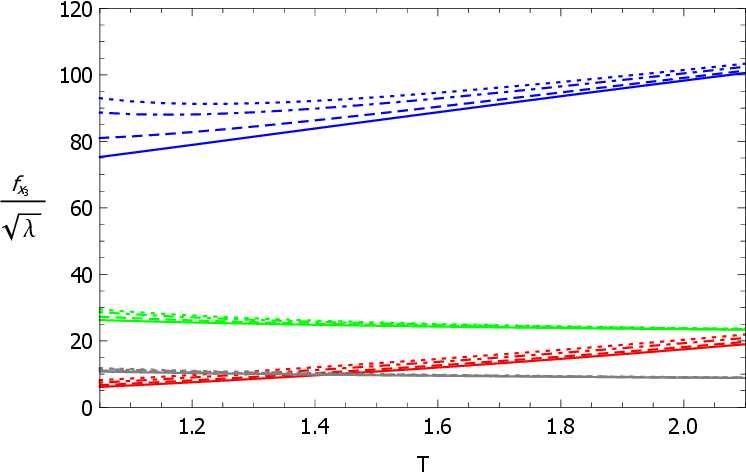}}
\caption{Directional drag forces $f_{x_i}/\sqrt{\lambda}$ as functions of the temperature $T$ at fixed $\mu=0.95$. The first and second rows correspond to $v=0.3$ and $0.9$, respectively. The red, blue, green, and gray curves denote $\nu=1$, $1.5$, $3$, and $4.5$, while the solid, dashed, dot-dashed, and dotted curves correspond to $c_B=0,-1,-3$, and $-5$. We use $R_{gg}=1.16\,\text{GeV}^2$, $p=0.273\,\text{GeV}^4$, and $q_3=5\,\text{GeV}^2$, with $[\mu]=[T]=[c_B]^{1/2}=[f_{x_i}]^{1/2}=\text{GeV}$.}
\label{figdrag5}
\end{figure*}

\begin{figure*}[t]
\centering
\subfigure[$f_{{x}_1}/{\sqrt{\lambda}}$ vs. $\mu$ for $v=0.3$]{
\label{figdrag6a}
\includegraphics[width=0.3\linewidth]{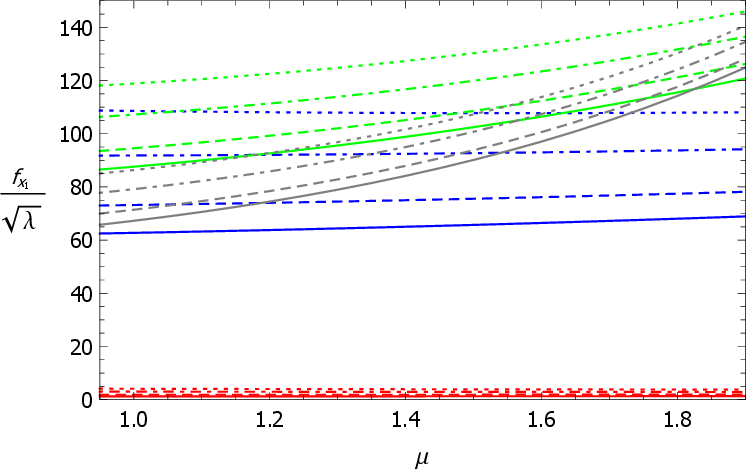}}
\subfigure[$f_{{x}_2}/{\sqrt{\lambda}}$ vs. $\mu$ for $v=0.3$]{
\label{figdrag6b}
\includegraphics[width=0.3\linewidth]{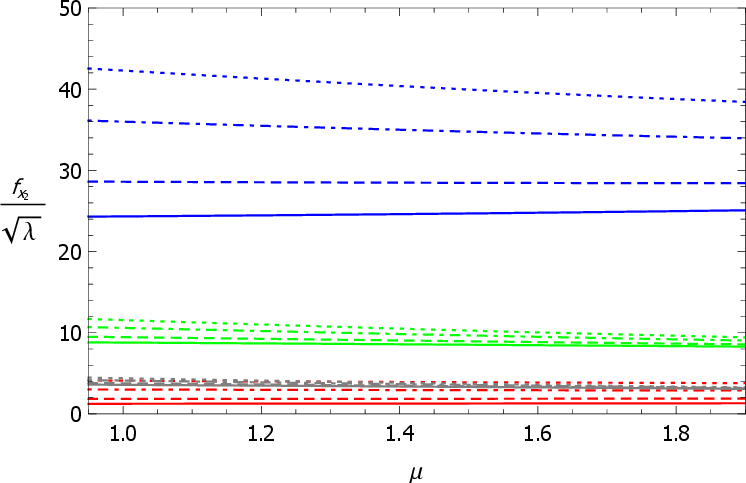}}
\subfigure[$f_{{x}_3}/{\sqrt{\lambda}}$ vs. $\mu$ for $v=0.3$]{
\label{figdrag6c}
\includegraphics[width=0.3\linewidth]{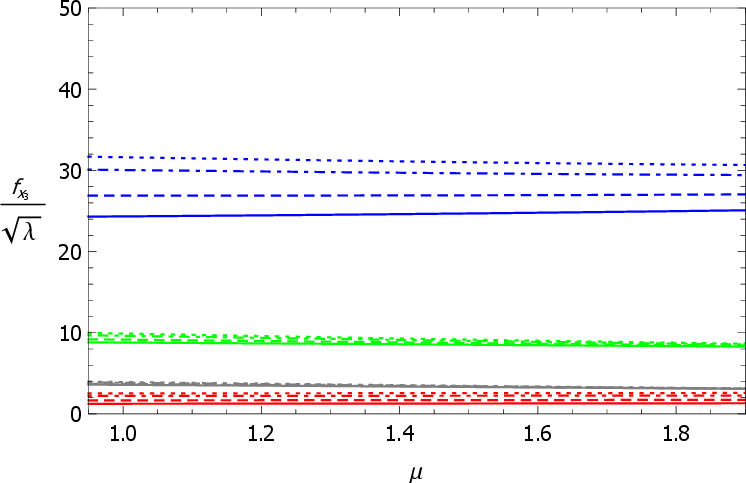}}
\subfigure[$f_{{x}_1}/{\sqrt{\lambda}}$ vs. $\mu$ for $v=0.9$]{
\label{figdrag6d}
\includegraphics[width=0.3\linewidth]{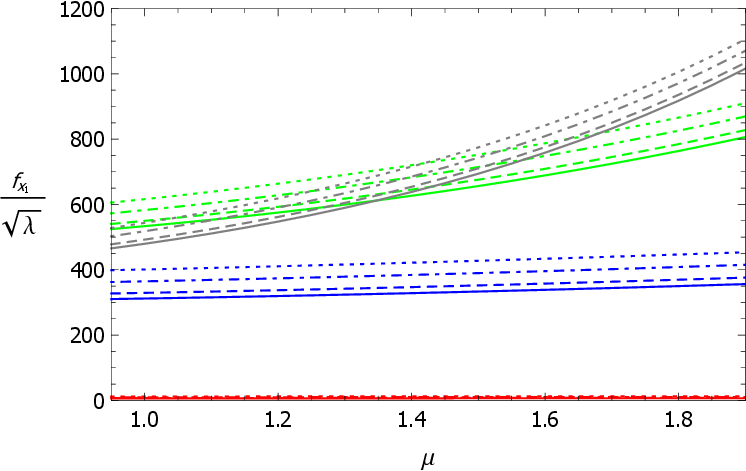}}
\subfigure[$f_{{x}_2}/{\sqrt{\lambda}}$ vs. $\mu$ for $v=0.9$]{
\label{figdrag6e}
\includegraphics[width=0.3\linewidth]{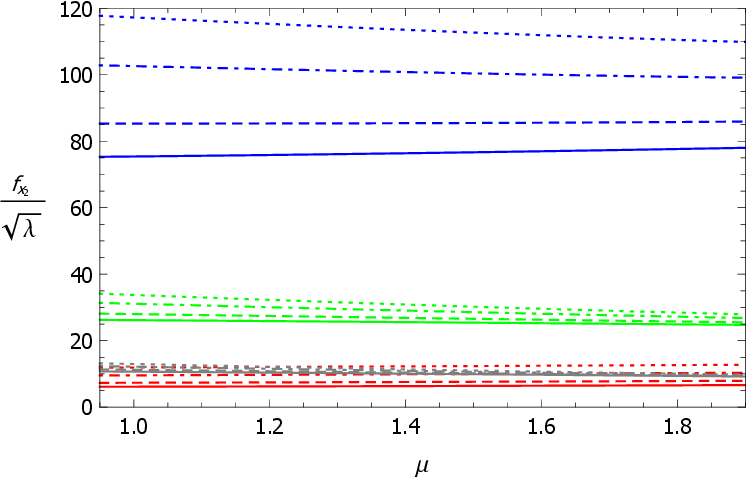}}
\subfigure[$f_{{x}_3}/{\sqrt{\lambda}}$ vs. $\mu$ for $v=0.9$]{
\label{figdrag6f}
\includegraphics[width=0.3\linewidth]{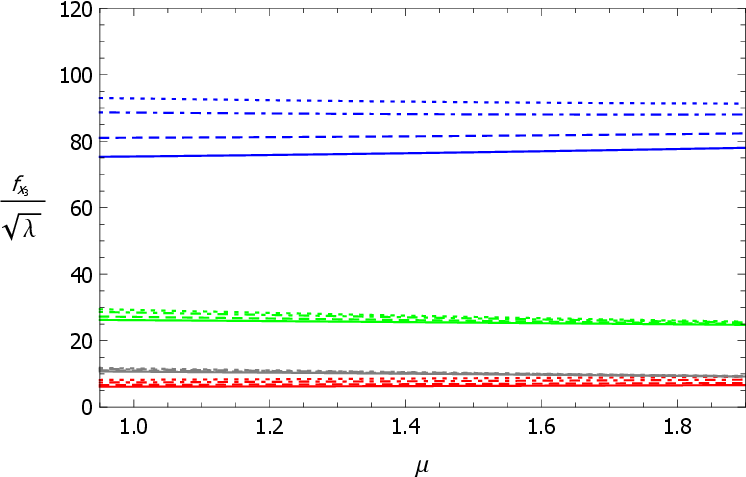}}
\caption{Directional drag forces $f_{x_i}/\sqrt{\lambda}$ as functions of the chemical potential $\mu$ at fixed $T=1.05$. The first and second rows correspond to $v=0.3$ and $0.9$, respectively. The red, blue, green, and gray curves denote $\nu=1$, $1.5$, $3$,  and $4.5$, while the solid, dashed, dot-dashed, and dotted curves correspond to $c_B=0,-1,-3$, and $-5$. We use $R_{gg}=1.16\,\text{GeV}^2$, $p=0.273\,\text{GeV}^4$, and $q_3=5\,\text{GeV}^2$, with $[\mu]=[T]=[c_B]^{1/2}=[f_{x_i}]^{1/2}=\text{GeV}$.}
\label{figdrag6}
\end{figure*}

Figures~\ref{figdrag4}--\ref{figdrag6} further display the dependence of $f_{x_i}/\sqrt{\lambda}$ on the probe velocity $v$, temperature $T$, and chemical potential $\mu$ for several fixed values of $c_B$ and $\nu$. In the isotropic QGP with magnetic catalysis, all three components grow rapidly with $v$. In the doubly anisotropic QGP, by contrast, only $f_{x_1}$ grows rapidly, whereas $f_{x_2}$ and $f_{x_3}$ are approximately linear in $v$. However, Figures~\ref{figdrag5} and~\ref{figdrag6} reveal that the temperature and chemical potential dependence of the drag force is modulated by other factors, including quark velocity, anisotropy, and magnetic field strength.

The main results for the drag force acting on a heavy-quark probe in the doubly anisotropic QGP with magnetic catalysis can be summarized as follows:
\begin{itemize}
\item Compared with that in the SYM system, the drag force on a heavy quark probe in an isotropic QGP without magnetic field is larger in magnitude. However, the ratio of the two drag forces decreases with increasing probe velocity and system temperature~\cite{Gursoy:2009kk}. Meanwhile, the isotropic QGP drag force itself increases monotonically with the chemical potential of the system~\cite{Caceres:2006dj, Chakraborty:2014kfa, Rougemont:2015wca, Xiong:2019wik, Heshmatian:2023yzz, Chen:2024epd}.

\item In the doubly anisotropic QGP with magnetic catalysis, the drag force anisotropy is governed mainly by the anisotropy parameter, making the parallel drag force significantly larger than the two transverse ones. Its effect is, however, modulated by temperature and chemical potential. For the parallel component, anisotropy always enhances the drag force relative to the isotropic reference; that is, the ratio of the parallel drag force to its isotropic reference value is greater than unity over the considered parameter range. The degree of this enhancement, however, depends on \(\nu\): at lower temperature and chemical potential, this ratio first increases and then decreases with increasing \(\nu\), whereas at higher temperature and chemical potential it increases monotonically with \(\nu\). For the two transverse components, within the considered temperature and chemical potential ranges, both first rise and then fall with increasing $\nu$. In particular, at strong anisotropy,  the two transverse drag forces at higher temperature and chemical potential are suppressed by $\nu$, whereas those at lower temperature and chemical potential remain enhanced. On the other hand, the introduction of an external magnetic field induces an anisotropy between the drag forces in the two transverse directions: specifically, the drag force along the magnetic field direction is smaller than that perpendicular to the magnetic field direction. Moreover, the external magnetic field strength enhances the drag forces in the three directions.

\item 
The anisotropy parameter does not change the trend of the parallel drag force with respect to quark velocity, but shifts the dependence of the two transverse drag forces on quark velocity from rapid growth in the isotropic case to approximately linear growth in the doubly anisotropic case. Moreover, the effects of temperature and chemical potential on the drag force may depend on various other factors, such as quark velocity, the anisotropy parameter, and the external magnetic field strength.
\end{itemize}

%In the following section, we will use this holographic doubly anisotropic background to analyze the chaotic origin of the drag force.

\section{The Chaotic Origin of the Drag Force}
\label{SEC:origin}
In previous sections, we examined the butterfly velocity and drag force within the holographic doubly anisotropic EMD framework. Here, we turn to the relation between them, namely the chaotic origin of the drag force~\cite{Ageev:2021poy}.
Reference~\cite{Ageev:2021poy} identified a relation between the drag force and butterfly velocity in $(d+1)$-dimensional isotropic holographic systems: for a slowly moving heavy-quark probe, the coefficient of the $v^3$ term in the near-horizon expansion of the properly normalized drag force is determined by $v_B^2$:
\begin{eqnarray}
f = - {\sigma} v -\mathcal{B}\, v^3 + \dots,\;\;\;\;\;\;\mathcal{B} = \frac{\sigma}{(d-1)v_B^2},
\label{dragexpiso46}
\end{eqnarray}
where $\sigma$ is the normalization coefficient and $\mathcal{B}$ is related to the butterfly velocity. In an anisotropic system, the relation remains valid when the drag force $f_{x_i}$ and butterfly velocity $v_{B,(i)}$ are evaluated along the same spatial direction and are connected through the following algebraic relations:
\begin{eqnarray}
f_{x_i} = - {\sigma}_i v -\mathcal{B}_i \, v^3 + \cdots,
\label{fxiexp}
\end{eqnarray}
and
\begin{eqnarray}
\label{vbalgebra}
v_{B, (i)} = \sqrt{\frac{\sigma_i^{-1}}{\displaystyle\sum_{j=1}^{d-1} \mathcal{B}_j  \sigma_j^{-2}}}.
\end{eqnarray}

However, recalling the previous derivation, the standard approach is in fact to compute the butterfly velocity via shock waves in the Einstein frame, while obtaining the drag force in the string frame. Therefore, when discussing the chaotic origin of the drag force, one must account for frame-dependent chaos, as proposed in Refs.~\cite{Shukla:2023pbp, Shukla:2024qlf, Lee:2025ias, Nongmaithem:2026rop}, which showed that quantum chaos (e.g., Poincaré sections and Lyapunov exponents) in the magnetized EMD gravity model was shown to lead to qualitatively opposite conclusions in the string and Einstein frames. 

Next, following the procedure for the chaotic origin of the drag force in Ref.~\cite{Ageev:2021poy}, we compute the directional butterfly velocity $v_{B,(i)}^{(S)}$ in the string frame for holographic anisotropic models. Within this derivation, $v_{B,(i)}^{(S)}$ characterizes quantum chaos in the string frame, as it is related to the coefficient in the expansion of $f_{x_i}$ in the quark velocity $v$. Then, via the Weyl transformation between the string and Einstein frames, we establish the relation between the drag force $f_{x_i}$ on a moving heavy quark in the doubly anisotropic QGP with magnetic catalysis and the butterfly velocity $v_{B,(i)}$ in the Einstein frame of this holographic system, thereby revealing the chaotic origin of the drag force while accounting for frame-dependent chaos.  

So as to expand the drag force according to Eq.~(\ref{fxiexp}),  we need to first expand $\tilde{G}_{tt}(z_c)$ and $\tilde{G}_{x_i x_i}(z_c)$ about the black-hole horizon:
\begin{eqnarray}
\tilde{G}_{tt}(z_c) &=& \tilde{G}_{t t}'(z_h) (z_c - z_h) + \cdots,\label{Taylorgtt}
\\
\tilde{G}_{x_i x_i}(z_c) &=& \tilde{G}_{x_i x_i}(z_h) + \tilde{G}_{x_i x_i}'(z_h) (z_c - z_h) + \cdots.\;\;\;\;\;
\label{Taylorgxx}
\end{eqnarray}
Solving Eq.~(\ref{sol zc}) near the horizon gives
\begin{eqnarray}
z_c = z_h - \frac{\tilde{G}_{x_i x_i}(z_h)}{\tilde{G}_{tt}'(z_h)}v^2+\cdots.
\label{Taylorzc}
\end{eqnarray}
Substituting Eqs.~(\ref{Taylorgtt}),   (\ref{Taylorgxx}) and~(\ref{Taylorzc}) into Eq.~(\ref{dragall}) yields the small-velocity expansion of the drag force along $x_i$:

\begin{eqnarray}
f_{x_i} &=& -\frac{\sqrt{\lambda}}{2\pi}\tilde{G}_{x_i x_i}(z_h)\,v\nonumber\\
&&-\frac{\sqrt{\lambda}}{2\pi}\left[-\tilde{G}_{x_i x_i}(z_h)\frac{\tilde{G}_{x_i x_i}'(z_h)}{\tilde{G}_{tt}'(z_h)}\right]v^3+\cdots.
\label{drag expand}
\end{eqnarray}
Comparing Eq.~(\ref{drag expand}) with Eq.~(\ref{fxiexp}) gives the coefficients $\sigma_i$ and $\mathcal{B}_i$:
\begin{eqnarray}
\label{sigmaandb}
\sigma_i &=& \frac{\sqrt{\lambda}}{2\pi}\tilde{G}_{x_i x_i}(z_h),\nonumber\\
\mathcal{B}_i &=& \frac{\sqrt{\lambda}}{2\pi}\left[-\tilde{G}_{x_i x_i}(z_h)\frac{\tilde{G}_{x_i x_i}'(z_h)}{\tilde{G}_{tt}'(z_h)}\right].
\label{coeff}
\end{eqnarray}
The directional butterfly velocity $v_{B,(i)}^{(S)}$ in the string frame is therefore
\begin{eqnarray}
v_{B,(i)}^{(S)} =\sqrt{\frac{- \tilde{G}_{tt}'(z_h)  \tilde{G}^{x_i x_i}(z_h) }{\displaystyle\sum_{j=1}^{3} \tilde{G}_{x_j x_j}'(z_h)  \tilde{G}^{x_j x_j}(z_h)}}.
\label{vbsecv}
\end{eqnarray}
The above analytical expressions are the butterfly velocities in the anisotropic holographic background obtained from the small \(v\) expansion of the drag force in the string frame. Note that the above expressions apply only to string-frame quantities. The corresponding butterfly velocity in the Einstein frame is given by Eq.~(\ref{vbappendix}). In particular, except for the form of the warp factor \(\mathfrak{b}_s(z)\) and  \(\mathfrak{b}(z)\) , all expressions in the string frame metric take the same form as those in the Einstein frame metric. 
As will be shown through the ratio \(\kappa\) below, the butterfly velocity in holographic anisotropic models is frame-dependent, consistent with Refs.~\cite{Shukla:2023pbp, Shukla:2024qlf, Lee:2025ias, Nongmaithem:2026rop}. A comparison of the analytical form of the warp factor \(\mathfrak{b}_s (z)\) shows that this frame dependence originates from the dilaton factor \(e^{2\phi(z)}\) and its gradient, which govern the Weyl transformation between the string and Einstein frames.

Using the Weyl transformation from the Einstein frame to the string frame $\tilde{G}_{\alpha \alpha} = e^{\sqrt{2/3}\phi}{G}_{\alpha \alpha}$ and imposing the black-hole horizon boundary condition \(g(z_h)=0\), the squared ratio of the butterfly velocities between the Einstein and string frames is given by
\begin{eqnarray}
\kappa= \frac{v_{B,(i)}^2}{\bigl(v_{B,(i)}^{(S)}\bigr)^2}
=1+
\frac{\sqrt{6}\,\phi'(z_h)}{\sum_{j=1}^{3}G^{x_j x_j}(z_h)G_{x_j x_j}'(z_h)}.
\label{vbvinv}
\end{eqnarray}
The relation for this squared ratio may hold universally for frame-dependent butterfly velocities, and the squared ratio itself is independent of directionality in holographic anisotropic backgrounds.
In particular, for holographic anisotropic models satisfying \(\phi'(z_h)=0\), the butterfly velocities in the Einstein and string frames coincide exactly. 
In this doubly anisotropic QGP with magnetic catalysis, the analytic squared ratio of the butterfly velocities between the Einstein and string frames is given by 
\begin{eqnarray}
\kappa= 1+\frac{\sqrt{6}\,\phi'(z_h)}
{2c_B z_h-\dfrac{2}{z_h}\left(1+\dfrac{2}{\nu}\right)+3\dfrac{\mathfrak b'(z_h)}{\mathfrak b(z_h)}}.
\label{kappathis}
\end{eqnarray}

\begin{figure*}[t]
\centering
\subfigure[$\kappa$ vs. $T$ for $\mu=0.95$]{
\label{figkappat1}
\includegraphics[width=0.4\linewidth]{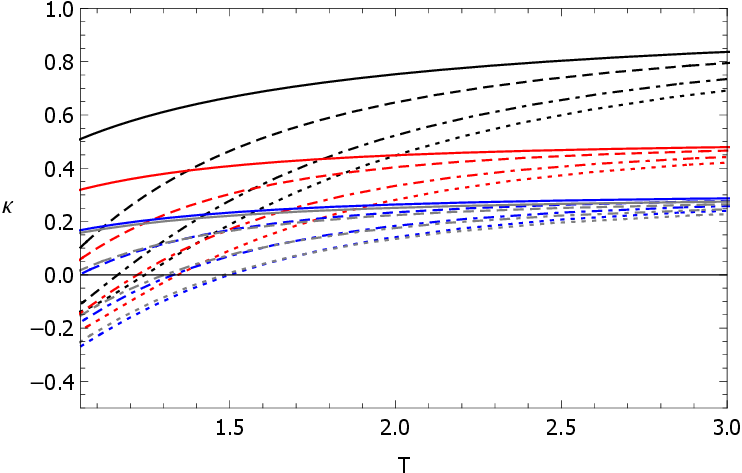}}
\subfigure[$\kappa$ vs. $T$ for $\mu=1.9$]{
\label{figkappat2}
\includegraphics[width=0.4\linewidth]{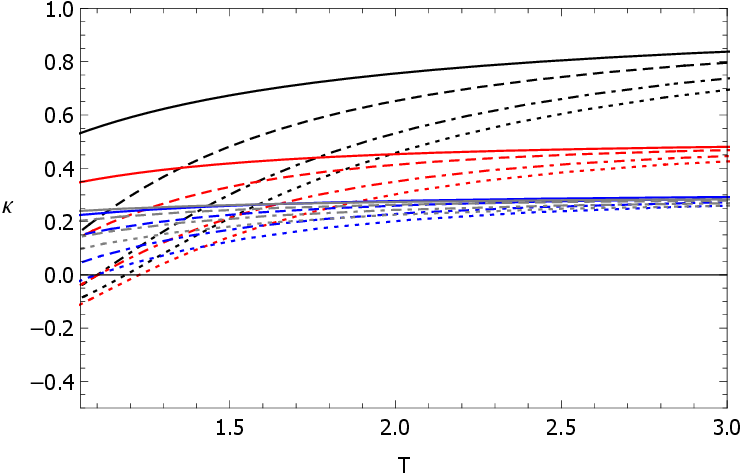}}
\subfigure[$\kappa$ vs. $\mu$ for $T=1.05$]{
\label{figkappamu1}
\includegraphics[width=0.4\linewidth]{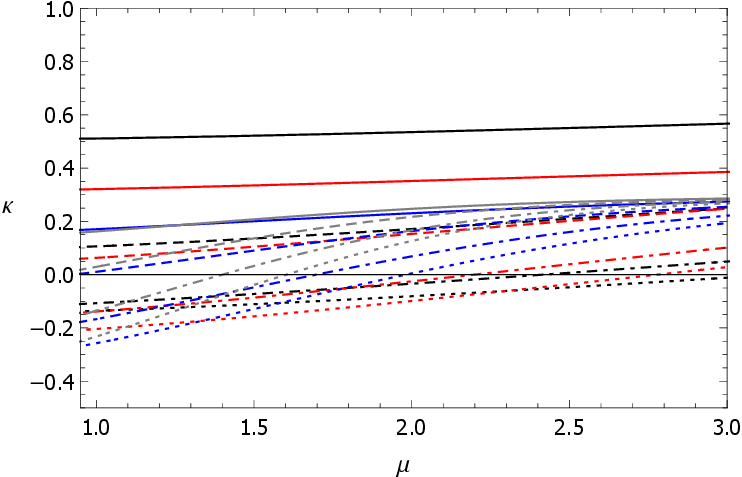}}
\subfigure[$\kappa$ vs. $\mu$ for $T=2.1$]{\label{figkappamu2}
\includegraphics[width=0.4\linewidth]{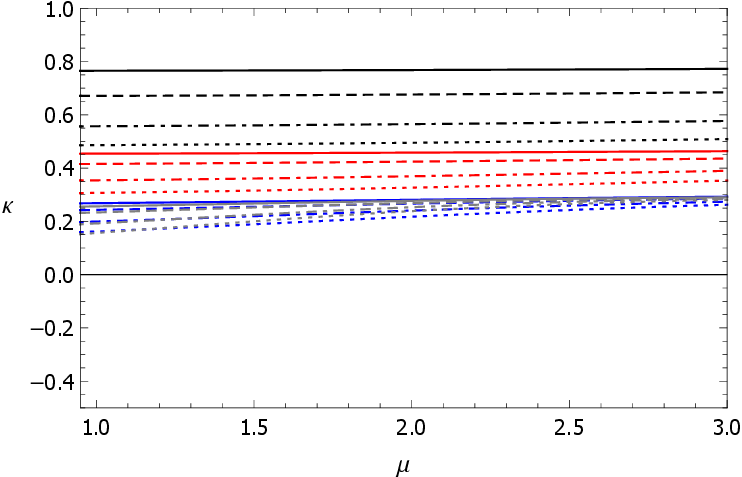}}
\caption{
The first row: \(\kappa\) versus \(T\) for \(\mu=0.95\) and \(1.9\). The second row: \(\kappa\) versus \(\mu\) for \(T=1.05\) and \(2.1\). The black, red, blue, and gray curves denote $\nu=1,1.5,3$, and $4.5$, while the solid, dashed, dot-dashed, and dotted curves correspond to $c_B=0,-1,-3$, and $-5$. We use $R_{gg}=1.16\,\text{GeV}^2$, $p=0.273\,\text{GeV}^4$, and $q_3=5\,\text{GeV}^2$, with $[\mu]=[T]=[c_B]^{1/2}=\text{GeV}$.}
\label{figkappa}
\end{figure*}

Figure~\ref{figkappa} shows that the squared ratio of the butterfly velocities between the Einstein and string frames increases with temperature and chemical potential. Note that $\kappa$ becomes negative at low temperature and chemical potential, implying unphysical results for the string-frame butterfly velocity $v_{B,(i)}^{(S)}$ in this region, whereas the Einstein-frame butterfly velocity \(v_{B,(i)}\) remains positive, as shown in Section~\ref{secBV}.

We now turn to discussing the chaotic origin of the drag force using the analytical expressions for the butterfly velocity in the Einstein frame.
we revisit the relation between the drag force and quantum chaos.
We first consider the isotropic limit, $\nu=1$ and $c_B=0$. Following Eq.~(\ref{dragexpiso46}), we define the normalized drag force as
\begin{eqnarray}
\label{drag57}
f_{\sigma,{{iso}}} = \frac{1}{\sigma} f_{{{iso}}} &=& - v - \frac{1}{3 [v^{(S)}_{B,({iso})}]^2} v^3.
\end{eqnarray}
Here the normalization factor is $\sigma=(\sqrt{\lambda}/2\pi)z_h^{-2}\mathfrak{b}_s(z_h)$.
Although the relation between the normalized drag force and the butterfly velocity in the string frame is formally independent of the explicit background geometry, the normalization factor, drag force, and butterfly velocity each retain their own geometric dependence. 
Then, substituting the definition of the frame-dependent butterfly velocity ratio $\kappa$, we identify the chaotic origin of the drag force while accounting for frame-dependent chaos in the isotropic limit, 
\begin{eqnarray}
f_{\sigma,{{iso}}} = - v - \frac{\kappa}{3\, v^2_{B,({iso})}} v^3.
\label{drag58}
\end{eqnarray}

\begin{figure*}[t]
\centering
 \subfigure[$f_{{iso}}$ vs. $v$]{
\label{figvb_drag1a}
\includegraphics[width=0.40\linewidth]{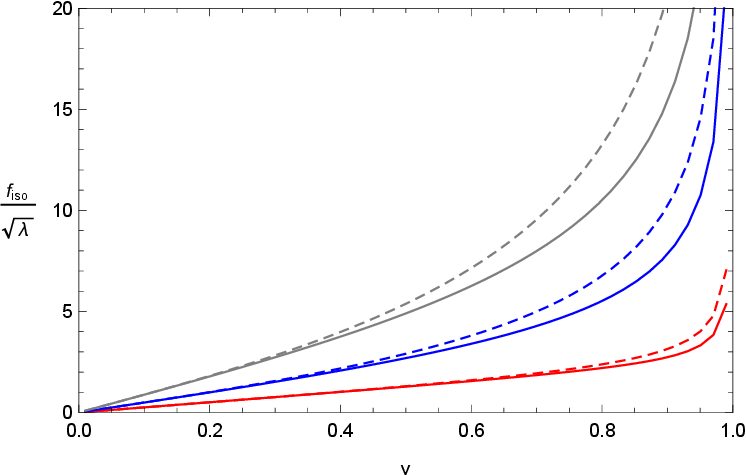}}
\subfigure[$f_{{iso}}/\sqrt{\lambda}$ vs. $v_{B,(iso)}$]{
\label{figvb_drag1b}
\includegraphics[width=0.40\linewidth]{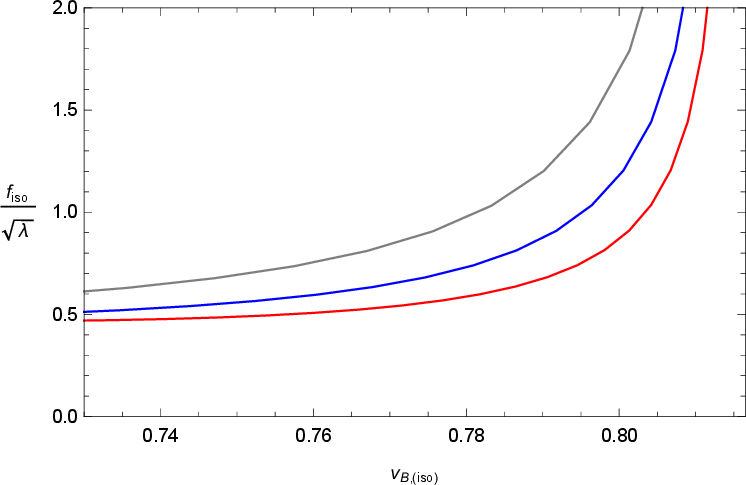}}
\caption{(a) Drag force obtained from Eq.~(\ref{drag22}) (solid curves) and from the small-velocity expression~(\ref{drag58}) (dashed curves). The red, blue, and gray curves correspond to $(\mu,T)=(0.8,0.6)$, $(1.6,1.2)$, and $(2.4,1.8)$, respectively. (b) Isotropic drag force in the chaos-related form as a function of the butterfly velocity in the Einstein frame for several values of $\mu$ at $v=0.2$. The red, blue, and gray curves correspond to $\mu=0.8$, $1.6$, and $2.4$, respectively. We use $R_{gg}=1.16\,\text{GeV}^2$, $p=0.273\,\text{GeV}^4$, and $q_3=5\,\text{GeV}^2$, with $[\mu]=[T]=[f]^{1/2}=\text{GeV}$.}
\label{figvb_drag1}
\end{figure*}

Figure~\ref{figvb_drag1a} compares the exact drag force with the small-velocity expression in Eq.~(\ref{drag58}). The two agree well approximately for \(v<0.35\). Figure~\ref{figvb_drag1b} shows that the drag force itself increases with $v_B$.
%, whereas Eq.~(\ref{drag58}) implies that the normalized drag force decreases as $v_B$ increases.

We next consider this relation in the doubly anisotropic QGP with magnetic catalysis. Eq.~(\ref{vbalgebra}) shows that the directional butterfly velocity $v_{B, (i)}$ is not determined by $\mathcal{B}_i$ and $\sigma_i$ alone; instead, all three spatial directions are coupled. In the present background, however, the directional butterfly velocities obey the chain of equalities in Eq.~(\ref{vbrelationship}).
This property yields an analytic relation between the normalized drag force along $x_i$ and the corresponding butterfly velocity:
\begin{eqnarray}
\label{andrag39}
f_{\sigma,{x_i}} = \frac{1}{\sigma_i} f_{x_i} = - v - \left(\frac{\kappa}{3\, v^2_{B,({i})}}  + \frac{\beta_i}{6\pi T}\right) v^3.
\end{eqnarray}
Here, the frame transformation of the butterfly velocity has already been carried out.
The normalized drag force in the doubly anisotropic QGP therefore differs from its isotropic counterpart by an additional correction at order $v^3$. We define the direction-dependent correction coefficient $\beta_i$ by
\begin{eqnarray}
\beta_1 &=& \frac{2(\nu-1)}{\nu z_h} + c_B z_h,
\\
\beta_2 &=& z_h^{1-\frac{2}{\nu}} \left( \frac{(1-\nu)}{\nu} + c_B z_h^2 \right),
\\
\beta_3 &=& z_h^{1-\frac{2}{\nu}} e^{c_B z_h^2} \left( \frac{(1-\nu)}{\nu} - 2c_B z_h^2 \right).
\end{eqnarray}
The coefficient $\beta_i$ encodes the directional dependence of the holographic geometry, while the normalization factor $\sigma_i$ is also direction dependent [see Eq.~(\ref{sigmaandb})].

\begin{figure*}[t]
\centering
\subfigure[$f_{x_1}$ vs. $v_{B, (1)}$ for $\mu=0.95$]{
\label{figvb_drag2a}
\includegraphics[width=0.3\linewidth]{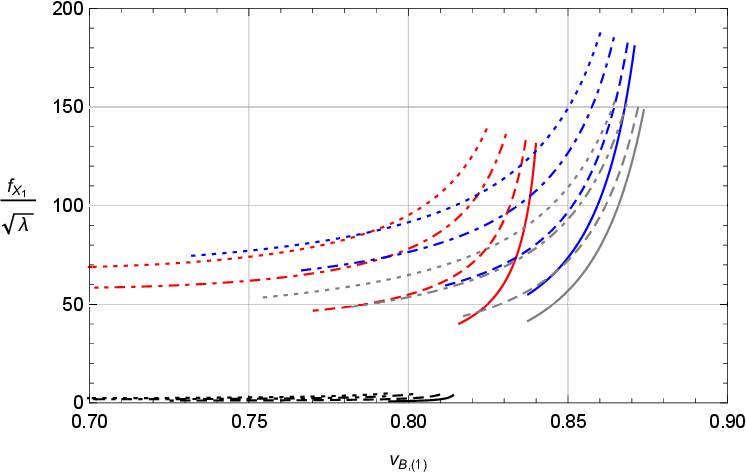}}
\subfigure[$f_{x_2}$ vs. $v_{B, (2)}$ for $\mu=0.95$]{
\label{figvb_drag2b}
\includegraphics[width=0.3\linewidth]{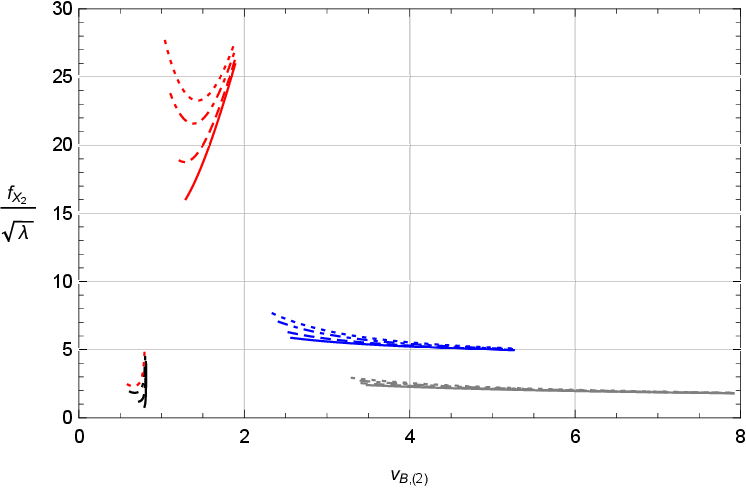}}
\subfigure[$f_{x_3}$ vs. $v_{B, (3)}$ for $\mu=0.95$]{
\label{figvb_drag2c}
\includegraphics[width=0.3\linewidth]{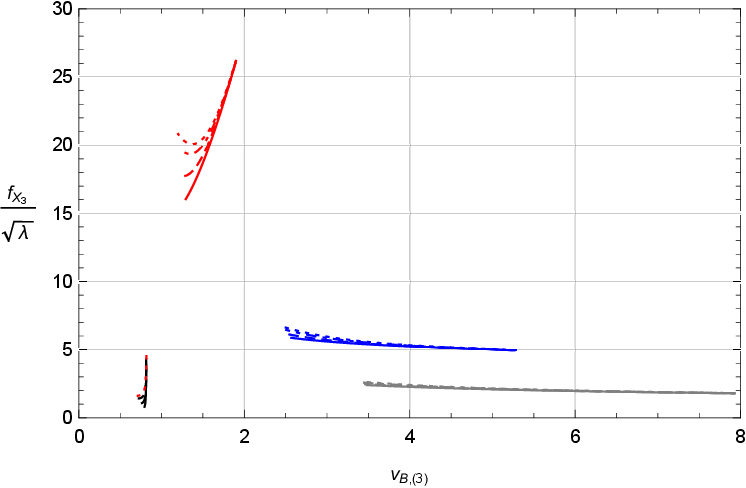}}
\subfigure[$f_{x_1}$ vs. $v_{B, (1)}$ for $\mu=1.9$]{
\label{figvb_drag2d}
\includegraphics[width=0.3\linewidth]{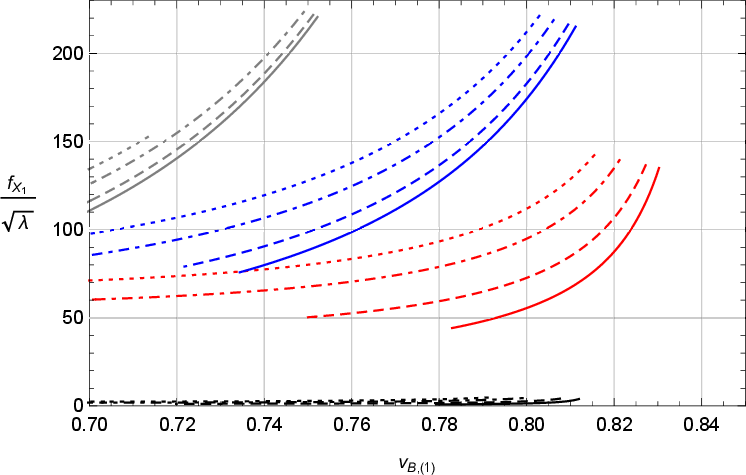}}
\subfigure[$f_{x_2}$ vs. $v_{B, (2)}$ for $\mu=1.9$]{
\label{figvb_drag2e}
\includegraphics[width=0.3\linewidth]{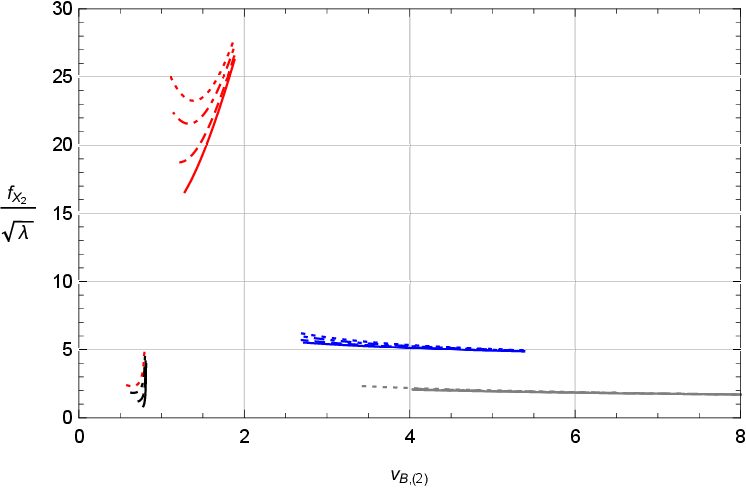}}
\subfigure[$f_{x_3}$ vs. $v_{B, (3)}$ for $\mu=1.9$]{
\label{figvb_drag2f}
\includegraphics[width=0.3\linewidth]{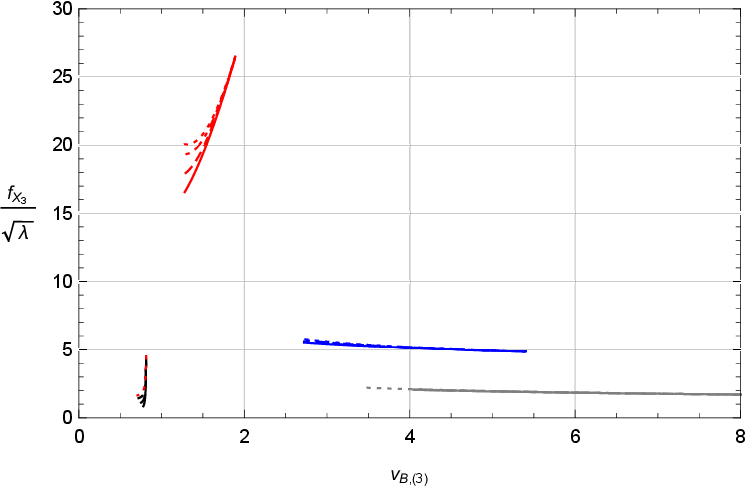}}
\caption{Directional drag forces $f_{x_i}$ in the chaos-related form as functions of $v_{B, (i)}$ at $v=0.2$. The first and second rows correspond to $\mu=0.95$ and $1.9$, respectively. The black, red, blue, and gray curves denote $\nu=1,1.5,3$, and $4.5$, while the solid, dashed, dot-dashed, and dotted curves correspond to $c_B=0,-1,-3$, and $-5$. We use $R_{gg}=1.16\,\text{GeV}^2$, $p=0.273\,\text{GeV}^4$, and $q_3=5\,\text{GeV}^2$, with $[\mu]=[T]=[c_B]^{1/2}=\text{GeV}$.}
\label{figvb_drag2}
\end{figure*}

Figure~\ref{figvb_drag2} shows the drag force in this chaos-related form for the doubly anisotropic QGP. We restrict our analysis to the pure deconfined QGP phase above the phase transition, with temperature \(T \in [1.05, 3.0]\), consistent with the results presented in Fig.~\ref{figvb4}. The horizontal range of the butterfly velocity varies significantly with the anisotropy parameter and magnetic field strength. Results show that the drag force in the primary anisotropic direction, \(f_{x_1}\), increases monotonically with the corresponding butterfly velocity \(v_{B, (1)}\), whereas the dependencies of the transverse drag forces on the corresponding butterfly velocities are more complex: under strong anisotropy, both \(f_{x_2}\) and \(f_{x_3}\) decrease with their respective butterfly velocities; under weak anisotropy and strong magnetic fields, they exhibit a non-monotonic trend, initially decreasing and then increasing.
Consistent with Sec.~\ref{sec:LOSS}, the magnetic field has only a weak effect on this relation, especially in the directions transverse to the primary anisotropy.

The main observations concerning the chaos-related form of the drag force are as follows:

\begin{itemize}
\item  In this doubly anisotropic holographic background, the chaotic origin of the drag force requires accounting for the frame dependence of the butterfly velocity. Moreover, the string-frame butterfly velocity may become unphysical.

\item In the doubly anisotropic QGP with magnetic catalysis, the analytic relation between the normalized drag force and butterfly velocity differs from the isotropic result by a direction-dependent correction at order $v^3$ that encodes the holographic geometry.

\end{itemize}

\section{Conclusion and discussion}
\label{sec:CON}

Using AdS/CFT duality, we have investigated heavy-quark energy loss and quantum chaos in a strongly coupled QGP with anisotropies generated by collision geometry and magnetic catalysis. Within the doubly anisotropic background of Ref.~\cite{Arefeva:2023jjh}, we have investigated three aspects: the directional butterfly velocities obtained from shock wave calculations in the Einstein frame, the directional drag forces on a moving heavy-quark probe in the string frame, and the frame-dependent relationship between the normalized drag force and butterfly velocity. Our numerical analysis was restricted to the pure deconfined phase and examined the effects of the primary anisotropy, magnetic field, temperature, chemical potential, and quark velocity. The resulting behavior differs in several respects from that of isotropic and singly anisotropic plasmas.

We studied the butterfly velocity, which characterizes the propagation of quantum information and is related to the drag force. In the isotropic QGP, it increases with temperature toward the conformal result and decreases with chemical potential.
In the anisotropic QGP with magnetic catalysis, the primary anisotropy is the dominant source of directional splitting in the butterfly velocity, whereas the magnetic field contributes much less. The two transverse components increase monotonically with $\nu$ and become superluminal under the coordinate normalization adopted here. This is an apparent effect that does not imply a violation of relativistic causality~\cite{Balasubramanian:2019stt, Basu:2025exh, Basu:2026wzp}. The dependence of the longitudinal component on $\nu$ is modulated by the chemical potential. The magnetic field has little effect on the transverse components but suppresses the longitudinal one. All three components increase monotonically with temperature, while their chemical-potential dependence is jointly controlled by the anisotropy and magnetic field. 

The directional dependence of the drag force reflects the anisotropic structure of the medium. In the isotropic QGP without magnetic catalysis, the ratio of the heavy-quark drag force to the $\mathcal{N}=4$ SYM result consistently exceeds unity, decreases monotonically with temperature, and increases monotonically with chemical potential. In the doubly anisotropic QGP, the anisotropy parameter dominates the directional dependence of the drag force, making the parallel component significantly larger than the transverse ones. However, this effect is heavily modulated by temperature and chemical potential, leading to regime-dependent monotonicities for the parallel component, and generally non-monotonic trends for the transverse components. Furthermore, the introduction of an external magnetic field breaks the symmetry between the two transverse directions, yielding a smaller drag force along the field direction than the perpendicular one, while generally enhancing the drag force in all three directions. Additionally, the anisotropy parameter changes the velocity dependence of the transverse drag forces from rapid growth in the isotropic case to approximately linear growth in the doubly anisotropic case. Ultimately, our findings reveal that the drag force is governed by a complex interplay among temperature, chemical potential, magnetic field, and quark velocity.

Finally, the frame-dependent normalized drag-chaos relation in the doubly anisotropic QGP differs from the isotropic result by a direction-dependent correction at order $v^3$ that encodes the holographic geometry. 
In the isotropic holographic background, the small-$v$ expansion of the drag force matches the exact result and increases with the butterfly velocity. Above the phase transition in the pure deconfined doubly anisotropic QGP, only the primary drag force increases monotonically with the butterfly velocity. The transverse drag forces, however, decrease monotonically under strong anisotropy, whereas under weak anisotropy and strong magnetic fields they first decrease and then increase.

\vspace{\baselineskip}

\section*{Acknowledgments}
We thank Xun Chen for discussions. This work was supported by the National Natural Science Foundation of China (NSFC) under Grants No. 12005056.

\bibliographystyle{apsrev4-2}
\bibliography{zladscft_refs}

%%%%%%%%%%%%%%%%%%%%%%%%%%%%%%%%%%%%%%%%%%%%%%%%%%%%%%%%%%%%%%%%%%%%%%%%%%%%%%%%
%%%%%%%%%%%%%%%%%%%%%%%%%%%%%%%%%%%%%%%%
\end{document}